\documentclass[%
 reprint,
 amsmath,amssymb,
 aps,
]{revtex4-2}
\usepackage{tabularx}
\usepackage{xcolor}
\usepackage{graphicx}
\usepackage{dcolumn}
\usepackage{bm}

\begin{document}

\preprint{APS/123-QED}

\title{ Magneto-optical transport in type-I multi-Weyl semimetals in the presence of orbital magnetic moment}


\author{Panchlal Prabhat}
\affiliation{ Department of Physics, Lalit Narayan Mithila University, Darbhanga, Bihar 846004, India}%

\author{Amit Gupta}
\email{sunnyamit31@gmail.com}
\affiliation{
  Department of Physics, M. R. M. College, Lalit Narayan Mithila University, Darbhanga, Bihar 846004, India}



\date{\today}

\begin{abstract}
Multi-Weyl semimetals (mWSMs) are topological quantum materials whose Weyl nodes carry higher monopole charges and exhibit strongly anisotropic dispersions. We investigate the linear and nonlinear magneto-optical transport of tilted type-I mWSMs with arbitrary monopole charge $J$ within a semiclassical Boltzmann framework that incorporates both Berry-curvature (BC) and orbital-magnetic-moment (OMM) corrections. We derive analytical expressions for the longitudinal, Hall, planar Hall and second-harmonic magnetoconductivities in the presence of external electric and magnetic fields. The resulting conductivity components exhibit characteristic scaling with the chemical potential and monopole charge, reflecting the anisotropic dispersion of multi-Weyl nodes. We further elucidate the interplay between the BC and OMM contributions and show that the OMM generally suppresses the BC-induced magnetoconductivity, while its contribution can become comparable to or larger than the BC contribution for specific transport geometries. Our results provide experimentally accessible signatures of the combined effects of cone tilt, higher monopole charge, Berry curvature, and orbital magnetic moment in the magneto-optical response of multi-Weyl semimetals.
\end{abstract}

\maketitle


\section{\label{sec:level1}
Introduction
 }
Topological quantum materials have emerged as a major research frontier in condensed-matter physics because their electronic properties are governed not only by band dispersion but also by the topology of Bloch wave functions. Weyl semimetals (WSMs) are a prominent class of three-dimensional topological phases in which the conduction and valence bands touch at isolated points, known as Weyl nodes, near the Fermi level \cite{murakami2007phase,wan2011topological,yang2011quantum,burkov2011weyl}. A Weyl node acts as a monopole or antimonopole of Berry curvature in momentum space and carries a quantized topological charge. Owing to the Nielsen Ninomiya theorem, Weyl nodes occur in pairs of opposite chirality in a periodic crystal, ensuring that the total topological charge in the Brillouin zone vanishes \cite{nielsen1981absence}. For an ordinary Weyl node, the topological charge is J=$\pm1$, associated with a linear dispersion in all three momentum directions and consequently with low-energy quasiparticles described by massless Weyl fermions. Weyl fermions and surface Fermi arcs have been experimentally established in several noncentrosymmetric materials, including the TaAs family, NbP, TaP, and NbAs, through angle-resolved photoemission and quantum-transport measurements \cite{xu2011chern,huang2015weyl,lv2015observation,xu2015discovery}.\\

A natural generalization is provided by multi-Weyl semimetals (mWSMs), whose nodes carry higher integer topological charges, J=2 or 3, protected by crystalline point-group symmetries \cite{fang2012multi,huang2016new,chen2016photonic,ahn2017optical}. In contrast to ordinary Weyl nodes, double- and triple-Weyl nodes exhibit quadratic and cubic dispersions, respectively, in the ($k_x)-(k_y$) plane while remaining linear along ($k_z$). The resulting anisotropic band structure and higher topological charge lead to unconventional electronic, thermoelectric, optical, and magnetotransport properties \cite{mai2017exploring,ahn2017optical,mukherjee2018doping,nag2020thermoelectric,nag2020magneto,menon2020anomalous,gupta2019novel,guptafloquet,gupta2022kerr}. Double-Weyl nodes have been proposed in compounds such as HgCr$_2$Se$_4$ and SrSi$_2$ and realized in photonic systems \cite{xu2011chern,huang2016new,chen2016photonic}. First-principles calculations have also identified quasi-one-dimensional molybdenum monochalcogenides ($A^I(MoX^{IV})_3$), with ($A^I$=Na,K,Rb,In,Ti) and ($X^{IV}$=$\mathrm{S,Se,Te}$), as candidate triple-Weyl systems \cite{liu2017predicted}.\\

Magnetotransport provides a sensitive probe of the Berry-geometric properties of Weyl systems. In the semiclassical wave-packet description, Berry curvature modifies the carrier velocity and phase-space dynamics, while the orbital magnetic moment (OMM), arising from the self-rotation of the Bloch wave packet, produces a magnetic-field-dependent correction to the quasiparticle energy \cite{sundaram1999wave}. These geometric effects can substantially modify the magnetoconductivity of WSMs. In particular, recent work has demonstrated that the OMM suppresses linear and nonlinear magnetoconductivities in tilted single-Weyl ((J=1)) systems \cite{gao2022suppression}. Nonlinear Hall transport in tilted multi-Weyl systems has also been investigated using lattice models without OMM corrections \cite{roy2022non}.  Moreover, in various planar Hall configurations, the OMM contribution has been found to be comparable to or even larger than the Berry-curvature contribution \cite{mandal2025anisotropic}. However, how the interplay between higher topological charge, cone tilt, Berry curvature and the OMM modifies the linear and nonlinear magnetoconductivities of multi-Weyl systems remains unexplored. This question is particularly relevant because the anisotropic dispersion of mWSMs changes both their density of states and Berry-geometric responses, potentially leading to scaling behavior that is qualitatively different from that of ordinary Weyl fermions.\\

In this work, we develop a semiclassical theory of magneto-optical transport in tilted type-I multi-Weyl semimetals with arbitrary topological charge (J). Using the Boltzmann transport equation in the relaxation-time approximation and consistently incorporating Berry-curvature and orbital-magnetic-moment corrections, we derive analytical expressions for the frequency-dependent longitudinal, Hall, planar Hall, and nonlinear magnetoconductivities. We identify the scaling of these responses with chemical potential and topological charge and clarify how cone tilt and the OMM modify the corresponding transport coefficients. We find that the OMM generally suppresses the Berry-curvature-induced magnetoconductivity in both linear and nonlinear response regimes while in some situations, the OMM-contributed parts turn out to be comparable to or even greater than the BC-only parts. Our results establish a unified framework for understanding the combined effects of topology, tilt and orbital magnetism in the magneto-optical response of multi-Weyl systems.\\

The remainder of the paper is organized as follows. Section \ref{model_ham} introduces the low-energy Hamiltonian for tilted type-I multi-Weyl semimetals and the semiclassical Boltzmann formalism with Berry-curvature and orbital-magnetic-moment corrections. Section \ref{linear_response} presents the linear magneto-optical response, including longitudinal, Hall, and planar Hall conductivities up to second order in the magnetic field. Section \ref{non-linear} discusses the second-order nonlinear magneto-optical response and second-harmonic generation. Section \ref{conclusion} summarizes our main results and conclusions. Technical details of the conductivity calculations are provided in Appendix A.\\

\section{Model Hamiltonian and Semiclassical Boltzmann approach}\label{model_ham}
The low-energy effective Hamiltonian for tilted m-WSMs with chirality s near a single Weyl point is given by \cite{mukherjee2018doping,nag2020thermoelectric,nag2020magneto,menon2020anomalous,gupta2019novel,guptafloquet,gupta2022kerr},
\begin{eqnarray}
\label{eq:ham}
\mathcal{H}_J=s\bigl[\alpha_J\hbar\{( k_{-})^J \sigma_{+}+ (k_{+})^J\sigma _{-}\}+\hbar \mathit{v}_F k_z\sigma _z \bigr]+\hbar \mathit{v}_F t_s k_z \sigma_{0}\nonumber\\
\end{eqnarray}
where $\sigma_{\pm}$=$\frac{1}{2}(\sigma_{x}\pm i\sigma_{y})$ and $k_{\pm}=k_x\pm i k_{y}$,  J represents topologcal charge, $\mathit{v}_F$ and $t_s$ are effective velocity and the tilt parameter along $\hat{z}$ direction respectively and  $\alpha_{J}$ is the material dependent parameter, e.g. $\alpha_1$ and $\alpha_2$ are the Fermi velocity and inverse of the mass respectively for the isotropic and double WSMs. The energy dispersion for tilted mWSMs is given by $ \epsilon_{\bm{ k}}^s=\hbar t_s \mathit{v}_F k_z+\chi\hbar\sqrt{\alpha_J^2 k_{\perp}^{2J}+\left(\mathit{v}_F k_z \right)^2}$ with $k_{\perp}=\sqrt{k_x^2+k_y^2}$ where the value +1(-1) for $\chi$ represents the conduction(valence) band. In this paper, we restrict ourselves to the type-I phases such that $\mid t_s\mid <$1, which gives a Fermi point or electron(hole) pocket, depending on whether the chemical potential cuts the nodal or conduction(valence) band. We will use semiclassical Boltzmann equations in this study.\\

In the presence of a static magnetic field $ \bm{B}$ and a time varying electric field $\bm {E}$, the semiclassical equations of motion at the location $\textbf{r}$ and the wave-vector  $\textbf{k}$ in a given band are \cite{sundaram1999wave, cortijo2016linear, zyuzin2017magnetotransport, zyuzin2017chiral, chang2015chiral, knoll2020negative, xiao2010berry}
\begin{eqnarray}
\bm{\dot{r}}=\frac{1}{\hbar}\nabla_{\bm k}\tilde{\varepsilon}_{\bm k}^{s}-\dot{\bm{k}}\times \bm\Omega_{\bm{k}}^s\label{EOMa}\\
\hbar \bm{\dot{k}}=-e\bm{E}-e\dot{\bm{r}}\times \bm{B}\label{EOMb}
\end{eqnarray}

\noindent where -e is the electron charge and $\Omega_{\bm{k}}^s$ is the Berry curvature(BC) \cite{sundaram1999wave,xiao2010berry}.  The first term on the right-hand side of Eq(\ref{EOMa}) is $\bm{\tilde{v}}_{ \bm{k}}^s = \frac{1}{\hbar} \nabla_{\bm k} \tilde{\varepsilon}_{\bm{ k}}^s$ =$\bm{v}_{ \bm{k}}^s-\frac{\partial }{\partial \bm k}(\bm{m}_{\bm{k}}^s\cdot \bm{B})$, defined in terms of an effective band dispersion $\tilde{\varepsilon}_{s}(\bm{ k})$ with $\bm{v}_{\bm{k}}^s =\frac{1}{\hbar} \nabla_{\bm k} \varepsilon_{\bm{ k}}^s$. In topological metals such as WSMs, this quantity acquires a term due to the intrinsic orbital moment,i.e., $\tilde{\epsilon_{\bm k}^s} = \epsilon_{\bm k}^s -\bm{m}_{\bm{k}}^s\cdot \bm{B}$,  while $\bm{m}_{\bm{k}}^s$ is the orbital magnetic moment(OMM) induced by the semiclassical “self-rotation” of the Bloch wave packet. The BC and OMM asociated with the $s^{th}$ band is given by \cite{sundaram1999wave,xiao2010berry}

\begin{eqnarray}
\bm{\Omega}_{\bm{k}}^s&=&Im[\langle \bm{\nabla}_k u_k^s\vert \times \vert \bm{\nabla}_k u_k^s\rangle]\\
\bm{m}_{\bm{k}}^s&=&-\frac{e}{2\hbar}Im[\langle \bm{\nabla}_k u_k^s\vert \times (\mathcal{H}_J(\bm k)- \epsilon_{\bm k}^s)\vert \bm{\nabla}_k u_k^s\rangle] 
\end{eqnarray}
where $\vert u_k^s\rangle$ denotes the periodic part of the Bloch wave function and satisfies the equation $ \mathcal{H}_J(\bm k)\vert u_k^s\rangle= \epsilon_{\bm k}^s\vert u_k^s\rangle $.\\

The general expressions for Berry curvature and orbital magnetic moment for multi-WSMs are \cite{dantas2018magnetotransport, nandy2021chiral}
\begin{eqnarray}
\bm{\Omega}_{\bm{k}}^s =- \frac{s}{2} \frac{J \mathit{v}_F \alpha_J^2 k_{\perp}^{2J-2}}{\beta_{\bm k,s}^3}\{k_x,k_y,J k_z\}\\
\bm{m}_{\bm{k}}^s=-\frac{s}{2} \frac{e J\mathit{v}_F \alpha_J^2 k_{\perp}^{2J-2}}{\beta_{\bm k,s}^2}\{k_x,k_y,J k_z
\}
\end{eqnarray}
with $\beta_{\bm k,s}=\sqrt{\alpha_J^2 k_{\perp}^{2J}+\left( k_z \mathit{v}_F \right)^2}$. \\

The two equations (\ref{EOMa}) and (\ref{EOMb}) can be decoupled to get

\begin{eqnarray}
\bm{\dot{r}}=\frac{1}{\hbar D}[\nabla_{\bm k}\tilde{\varepsilon}_{\bm k}^{s}+e\bm{E}\times\bm{\Omega}_{\bm{k}}^s
+\frac{e}{\hbar}(\nabla_{\bm k}\tilde{\varepsilon}_{\bm k}^{s}\cdot \bm{\Omega}_{\bm{k}}^s )\bm{B} ]\label{eomdecouplea}\\
\bm{\dot{k}}=\frac{1}{\hbar D}[-e\bm{E}-\frac{e}{\hbar}\nabla_{\bm k}\tilde{\varepsilon}_{\bm k}^{s}\times \bm{B}-\frac{e^2}{\hbar}(\bm{E}.\bm{B}) \bm{\Omega}_{\bm{k}}^s]\label{eomdecoupleb}
\end{eqnarray}
 
\noindent where the factor $D=1+\frac{e}{\hbar}( \bm{\Omega}_{\bm{k}}^s\cdot \bm{B})$ modifies the phase space volume \cite{xiao2005berry}. These equations contain three distinct contributions to carrier dynamics: $\bm{\tilde{v}}_{ \bm{k}}^s$ which governs ordinary charge transport in the absence of Berry curvature. Anomalous velocity, proportional to $\bm{E}\times\bm{\Omega}_{\bm{k}}^s$ which produces intrinsic Hall-type responses even without Lorentz-force deflection. Magnetic-field correction, arising from $(\bm{v}_{ \bm{k}}^s\cdot \bm{\Omega}_{\bm{k}}^s )\bm{B}$ which reflects the chiral nature of Weyl fermions and plays an essential role in magnetoconductivity.\\

For a given chirality $s =\pm $ of a multi-Weyl node, the semiclassical Boltzmann equation (SBE) in the relaxation time approximation(RTA) reads as follows\\

\begin{eqnarray}
\frac{\partial \tilde{f}^s}{\partial t}+\bm{\dot{k}}.\frac{\partial \tilde{f}^s}{\partial \bm k}=\frac{ \tilde{f}^s-\tilde{f}^s_0}{\tau}\label{SBE}
\end{eqnarray}

Here, $\tilde{f}^s(\tilde{\epsilon_{\bm k}^s})$ is the electron distribution function and $\tau$ is the relaxation time originating from the scattering of electrons by phonons, impurities, electrons, and other lattice imperfections \cite{malic2011microscopic}. We treat $\tau$ as a phenomenological constant and consider intranode scattering only. \\

The $\tilde{f}_{0}^{s}( \epsilon_{\bm k}^s)$ can be expanded at low magnetic field as
\cite{pellegrino2015helicons}

\begin{eqnarray}
\tilde{f}_{0}^{s}(\tilde{\epsilon_{\bm k}^s})=\tilde{f}_{0}^{s}( \epsilon_{\bm k}^s -\bm{m}_{\bm{k}}^s\cdot \bm{B})\nonumber\\
\simeq \tilde{f}_{0}^{s}( \epsilon_{\bm k}^s)-\bm{m}_{\bm{k}}^s\cdot \bm{B}\frac{\partial \tilde{f}_{0}^{s}( \epsilon_{\bm k}^s)}{\partial \epsilon_{\bm k}^s}
\end{eqnarray}

\noindent where $\tilde{f}_{0}^{s}( \epsilon_{\bm k}^s)=1/[e^{ (\epsilon_{\bm k}^s-\mu)/k_B T} +1]$ with $k_B$ the Boltzmann constant, T the temperature, and $\mu$ the chemical
potential. Eq.(\ref{SBE}) can be solved by expanding the distribution function as a power series in the electric field as 

\begin{equation}
\tilde{f}^{s}=\tilde{f}_{0}^{s}+\tilde{f}_{1}^{s} e^{-i\omega t}+\tilde{f}_{2}^{s}e^{-2i\omega t}+....\label{fexp}
\end{equation}

\noindent where $\tilde{f}_{1}^{s}$ and $\tilde{f}_{2}^{s}$ are the first- and second-order terms for $\bm{E}$, respectively. The electric current density can be calculated by

\begin{eqnarray}
\bm{j}=-\frac{e}{(2\pi)^3}\int d^3k D\bm{\dot r}\tilde{f}^{s} \label{curr_den_main}
\end{eqnarray}

Equation(\ref{curr_den_main}) computes the conductivity components under the combined action of external electric and magnetic fields. The J=1 case has been discussed by the authors of Ref.\cite{gao2022suppression}. In Table 1, we have summarized various measurable quantities between multi-WSMs derived from the explicit expressions in the sections that follow .\\ 

\section{Linear response of multi-WSMs}\label{linear_response}
For linear electric field response, we retain only the first two terms of Eq.(\ref{fexp}) and substitute Eq.(\ref{eomdecoupleb}) in Eq.(\ref{SBE})
\begin{eqnarray}
\frac{1}{\hbar D}[-e \textbf{E}-\frac{e^2}{\hbar}(\textbf{E}\cdot\textbf{B})\bm{\Omega}_{\bm{k}}^s]\cdot \frac{\partial \tilde{f}_0^s}{\partial \bm k} -i \omega \tilde{f}_1^s=-\frac{\tilde{f}_1^s}{\tau}
\end{eqnarray}
Solving for $\tilde{f}_{1}^{s}$, we obtain the following
\begin{equation}
\tilde{f}_1^s=\frac{\tau}{(1-i\omega \tau)}\frac{1}{\hbar D}[e \textbf{E}+\frac{e^2}{\hbar}(\textbf{E}\cdot\textbf{B})\bm \Omega_{\bm k}^s]\cdot \frac{\partial \tilde{f}_0^s}{\partial \bm k}\label{linearE}
\end{equation}

We expand Eq.(\ref{linearE}) up to the second order in the magnetic field and obtain
\begin{widetext}
\begin{eqnarray}\label{disf}
\tilde{f}_1^s&=&\frac{\tau}{(1-i\omega \tau)}\biggl[e \textbf{E}\cdot \bm{\mathit{v}}_{\bm k}^s\cdot \frac{\partial \tilde{f}_0^s}{\partial \epsilon_{\bm k}^s}-\frac{e^2}{\hbar}(\textbf{B}\cdot \bm\Omega_{\bm k}^s)(\textbf{E}\cdot \bm{\mathit{v}}_{\bm k}^s )\frac{\partial \tilde{f}_0^s}{\partial \epsilon_{\bm k}^s}+\frac{e^2}{\hbar}(\textbf{E}\cdot \textbf{B})(\bm\Omega_{\bm k}^s \cdot \bm{\mathit{v}}_{\bm k}^s )\frac{\partial \tilde{f}_0^s}{\partial \epsilon_{\bm k}^s}-\frac{e}{\hbar}\textbf{E}\cdot\frac{\partial}{\partial \bm k}\biggl(\textbf{m}_{\bm k}^s\cdot \textbf{B}\frac{\partial \tilde{f}_0^s}{\partial \epsilon_{\bm k}^s}\biggr)\nonumber\\
&-&\frac{e^3}{\hbar^2}(\textbf{B}\cdot \bm\Omega_{\bm k}^s)(\textbf{E}\cdot \textbf{B})(\bm\Omega_{\bm k}^s \cdot \bm{\mathit{v}}_{\bm k}^s )\frac{\partial \tilde{f}_0^s}{\partial \epsilon_{\bm k}^s}+\frac{e^3}{\hbar^2}(\textbf{B}\cdot \bm\Omega_{\bm k}^s)^2(\textbf{E}\cdot \bm{\mathit{v}}_{\bm k}^s )\frac{\partial \tilde{f}_0^s}{\partial \epsilon_{\bm k}^s}+\frac{e^2}{\hbar^2}(\textbf{B}\cdot \bm\Omega_{\bm k}^s)\textbf{E}\cdot \frac{\partial}{\partial \bm k}\biggl(\textbf{m}_{\bm k}^s\cdot \textbf{B}\frac{\partial \tilde{f}_0^s}{\partial \epsilon_{\bm k}^s}\biggr)\nonumber\\
&-&\frac{e^2}{\hbar^2}\bm\Omega_{\bm k}^s\cdot \frac{\partial}{\partial \bm k}\biggl(\textbf{m}_{\bm k}^s\cdot \textbf{B}\frac{\partial \tilde{f}_0^s}{\partial \epsilon_{\bm k}^s}\biggr)\biggr]
\end{eqnarray}
\end{widetext}

From Eqs.(\ref{eomdecouplea}) and Eq.(\ref{disf}), the expression for current density at time t is given by
\begin{eqnarray}
\bm{j}_1&=&-\frac{e}{(2\pi)^3}\int d^3k \Bigl[\bm{\mathit{\tilde{v}}}_{\bm k}^s+\frac{e}{\hbar}(\bm\Omega_{\bm k}^s  \cdot \bm{\mathit{\tilde{v}}}_{\bm k}^s )\textbf{B}\Bigr]\tilde{f}_1^{s}\label{cur_den} \nonumber\\
&-&\frac{e^2}{2\pi)^3\hbar}\int d^3k \textbf{E}\times \bm\Omega_{\bm k}^s \tilde{f}_0^s\label{cond_eqn}
\end{eqnarray}

The above equation can be expressed in frequency space $\omega$ as
\begin{equation}
j_a(\omega)=\sigma_{ab}(\omega)E_b(\omega)
\end{equation}

\noindent where $\sigma_{ab}(\omega)$ is the frequency dependent conductivity. The typical longitudinal or Hall conductivities are known to be formed by a single contribution from the group velocity $\bm{\mathit{\tilde{v}}}_{\bm k}^s$  or the Berry curvature $\bm{\Omega}_{\bm{k}}^s$. In addition to the standard components, the coupling terms between the group velocity $\bm{\mathit{\tilde{v}}}_{\bm k}^s$ and the Berry curvature  $\bm{\Omega}_{\bm{k}}^s$ cause the conductivity $\sigma_{ab}(\omega)$  to arise in the presence of the magnetic field  [see Eq. (\ref{cond_eqn})]. The external magnetic field triggers their combined contributions, which are essential to the electron transit.\\

\subsection{Calculations of conductivities components without magnetic field}
In the absence of a magnetic field, the optical conductivity is determined solely by the band
velocity and equilibrium electronic distribution. Because multi-Weyl semimetals possess
highly anisotropic electronic dispersions, the conductivity tensor is also anisotropic.
Substituting Eq.(\ref{disf}) without $ \textbf{B} $ terms into the first term of Eq.(\ref{cur_den}), we get diagonal conductivities components for single Weyl nodes
\begin{equation}
\sigma_{ab}^{(0)}(\omega)=\frac{\tau}{(1-i\omega \tau)}\frac{e^2}{(2\pi)^3}\int d^3k \mathit{v}_a^s \mathit{v}_b^s\Bigl(-\frac{\partial f_0^s}{\partial \epsilon_{\bm k}^s}\Bigr)
\end{equation}

At T=0 K, we use $-\frac{\partial f_0^s}{\partial \varepsilon_{\bm k}^s}=\delta(\varepsilon_{\bm k}^s-\mu)$ to get

\begin{widetext}
\begin{eqnarray}
\sigma_{zz}^{(0)}(\omega)=\frac{\mu^{\frac{2}{J}}\mathit{v_F}\hbar^{-\frac{2}{J}}\Gamma(1/J)(1-t_s)^{(1-\frac{1}{J})}(1+t_s)^{-(1+\frac{1}{J})}\sigma_D}{\pi^{\frac{1}{2}}\alpha_J^{2/J}\Gamma(1/2+1/J)}\Bigl[1-\frac{2(J+1)}{(J+2)}\frac{(1+t_s)^{(1+\frac{1}{J})}}{(1-t_s)^{(2+\frac{1}{J})}} {}_2F_1\Bigl(2+\frac{1}{J},\frac{2+J}{J},2+\frac{2}{J},\frac{2t_s}{-1+t_s}\Bigr)\Bigr]\label{cond_noB_triple_zz}\nonumber\\
\end{eqnarray}
\begin{eqnarray}
\sigma_{xx}^{(0)}(\omega)=\sigma_{yy}^0(\omega)=\frac{J\mu^2\sigma_D}{\pi\mathit{v_F}\hbar^2t_s^3}\Bigl[\frac{t_s}{(1-t_s^2)}+\frac{1}{2}\ln\frac{1-t_s}{1+t_s}\Bigr]\label{cond_noB_triple_xx}
\end{eqnarray}
\end{widetext}

\noindent where $\sigma_D=\frac{e^2 \tau}{(1-i\omega \tau)4\pi \hbar}$ is frequency dependent complex Drude conductivity and $_2F_1{(a,b,c,z)}=\frac{\Gamma(c)}{\Gamma(b)\Gamma(c-b)}\int_0^1 dt \frac{t^{b-1}(1-t)}{(1-t z)^a}$ is the hypergeometric function \cite{arfken2011mathematical}. Equations (\ref{cond_noB_triple_zz}) and (\ref{cond_noB_triple_xx}) are consistent with the results of Refs.\cite{ghosh2024direction}. These conductivity components are independent of chirality and are even functions of the tilt parameter $t_s$. Consequently, the total conductivity is twice the contribution from a single multi-Weyl node. The $\sigma_{zz}^{(0)}(\omega)$ components exhibit different tilt dependence   with J while $\sigma_{xx}^{(0)}(\omega)$ have the same tilt dependence. 
This difference reflects the anisotropic dispersion of multi-Weyl fermions in the out-of-plane and in-plane momentum directions. In the limit $t\rightarrow$0, $\sigma_{xx}^{(0)}(\omega)=\sigma_{yy}^{(0)}(\omega)=\frac{2J\mu^2}{3\pi \hbar^2 v_F}\sigma_D$, $\sigma_{zz}^{(0)}(\omega)=\frac{v_F\mu^{2/J}\Gamma(1+1/J)}{2\pi^{1/2}\alpha_J^{2/J}\Gamma(3/2+1/J)}\sigma_D$. Notably, the conductivity components obey distinct analytic scaling laws with the chemical potential and topological charge, namely $\sigma_{zz}^{(0)}(\omega)\propto v_F\mu^{2/J}/\alpha_J^{2/J}$ and $\sigma_{xx}^{(0)}(\omega)\propto J\mu^2$. The tilt dependence of $\sigma_{xx}^{(0)}(\omega)$ and $\sigma_{zz}^{(0)}(\omega)$ for multi-Weyl semimetals is presented in Fig.~(\ref{fig_cond_noB_tilt}a). Figure~(\ref{fig_cond_noB_tilt}b) further shows that the frequency dependence of $\sigma_{aa}^{(0)}(\omega)$ ($a=x,z$) exhibits the characteristic Drude response. Notably, for $J>1$, the $\sigma_{zz}^{(0)}(\omega)$ conductivity is consistently larger than $\sigma_{xx}^{(0)}(\omega)$ one. The conductivity elements are estimated at $\omega \rightarrow 0$ in Table(\ref{tab:table3}).\\

\subsection{Calculations of conductivity components linear in magnetic field}
Substituting Eq.(\ref{disf}) with $ \textbf{B} $ terms up to first order into the first term of Eq.(\ref{cur_den}), we get the conductivities components
\begin{equation}\label{cond_B} 
\sigma_{ab}^{(B)}(\omega)=\sigma_{ab}^{(B,\Omega)}(\omega)+\sigma_{ab}^{(B,m)}(\omega)
\end{equation}
where
\begin{widetext}
\begin{eqnarray}
\sigma_{ab}^{(B,\Omega)}(\omega)&=&\frac{\tau}{\hbar(1-i\omega \tau)}\frac{e^3}{(2\pi)^3}\int d^3k [(\mathit{v}_a^s B_b+\mathit{v}_b^s B_a)(\bm\Omega_{\bm k}^s \cdot \bm{\mathit{v}}_{\bm k}^s )-\mathit{v}_a^s \mathit{v}_b^s(\bm\Omega_{\bm k}^s \cdot \textbf{B})]\Bigl(-\frac{\partial f_0^s}{\partial \epsilon_{\bm k}^s}\Bigr)\label{cond_B_Omega}\\
\sigma_{ab}^{(B,m)}(\omega)&=&\frac{\tau}{\hbar(1-i\omega \tau)}\frac{e^2}{(2\pi)^3}\int d^3k\Bigl[\frac{\partial \mathit{v}_a^s }{\partial k_b}(\textbf{m}_{\bm k}^s\cdot \textbf{B})-\frac{\partial (\textbf{m}_{\bm k}^s\cdot \textbf{B})}{\partial k_a}\mathit{v}_b^s \Bigr]\Bigl(-\frac{\partial f_0^s}{\partial \epsilon_{\bm k}^s}\Bigr)\label{condcomp}
\end{eqnarray}
\end{widetext}

We can easily check from Eqs.(\ref{cond_B}) that $\sigma_{ab}^{(B)}(\omega)=\sigma_{ba}^{(B)}(\omega)$. Without a tilt term, the system possesses time-reversal symmetry; $\sigma_{ab}^{(B)}$ needs to obey the Onsager relation, resulting in zero conductivity \cite{onsager1931reciprocal, morimoto2016semiclassical}. However, the time-reversal symmetry is destroyed for a finite value of tilt $t_s\neq0$. The first term and the second term in Eq. (\ref{cond_B}) correspond to the Berry curvature $\Omega_{\bm k}$ and
the orbital magnetic moment $m_{\bm k}$ respectively. The following section provides a detailed analysis of Eq.(\ref{cond_B}) using the magnetic field perpendicular and parallel to the tilt direction $ \hat{\bm t}_s$.\\

\onecolumngrid\
\begin{center}\
\begin{figure}
\begin{tabular}{ccc}
\includegraphics[width=0.5\linewidth]{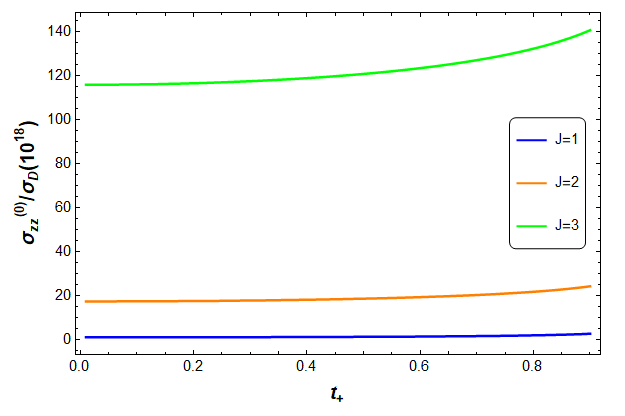} &
\includegraphics[width=0.5\linewidth]{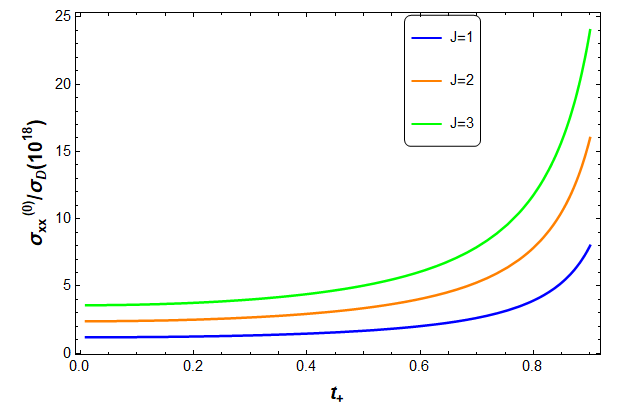} \\
\multicolumn{2}{c}{(a)}\\
\includegraphics[width=0.5\linewidth]{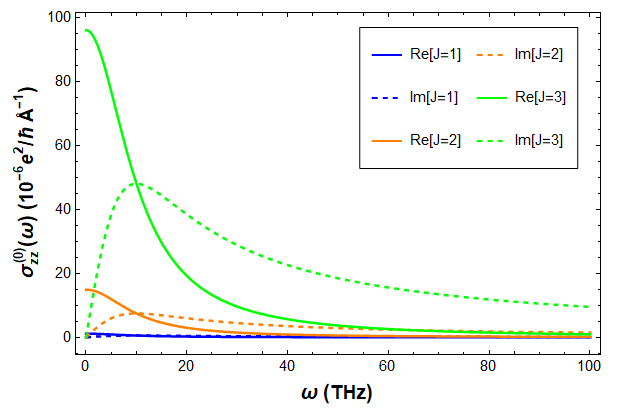} &
\includegraphics[width=0.5\linewidth]{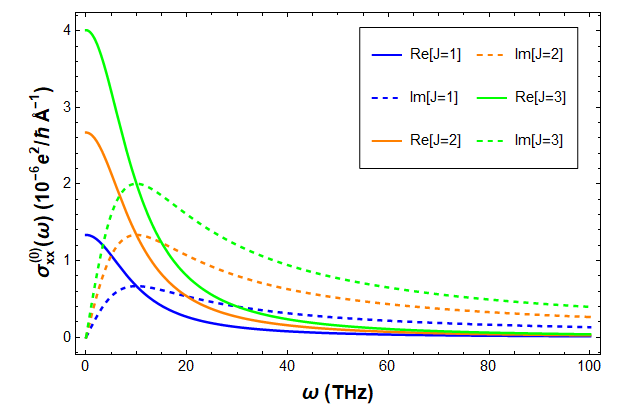} \\
\multicolumn{2}{c}{(b)}\\
\end{tabular}
\caption{(a) The dependence of the optical conductivity $\sigma^{(0)}_{zz}$ and $\sigma^{(0)}_{xx}$ on the tilt $t_+$ at zero B field. (b) The frequency dependence of the Drude conductivity at $t_s=0.5$. The other parameters are taken as $v_F=4.13\times 10^5$ m / s, $\alpha_2=0.009 m^2/s$, $\alpha_3=4.5\times 10^{-11} m^3/s$, $\mu=1 meV$ and $\tau=10^{-13}s$. }
\label{fig_cond_noB_tilt}\
\end{figure}\
\end{center}\
\twocolumngrid\

\onecolumngrid\
\begin{center}\
\begin{figure}
\begin{tabular}{ccc}
\includegraphics[width=0.35\linewidth]{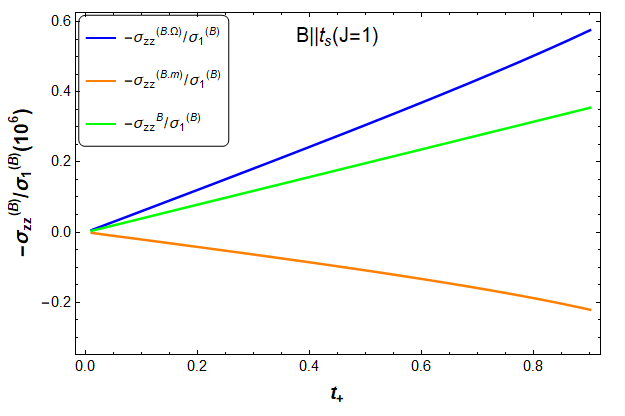} &
\includegraphics[width=0.35\linewidth]{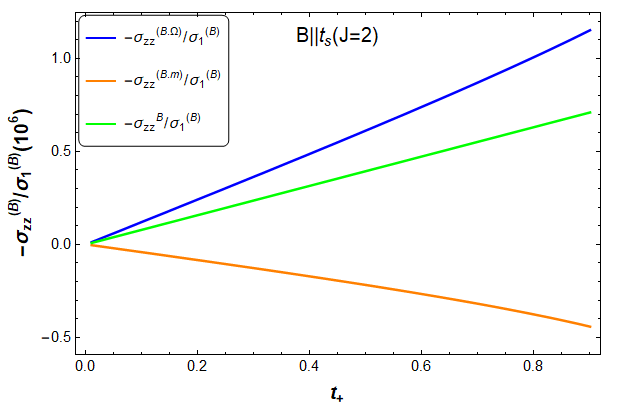} &
\includegraphics[width=0.35\linewidth]{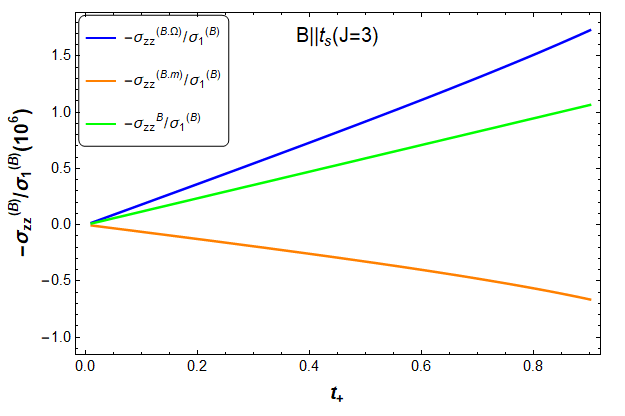} \\ 
\multicolumn{3}{c}{(a)}\\
\includegraphics[width=0.35\linewidth]{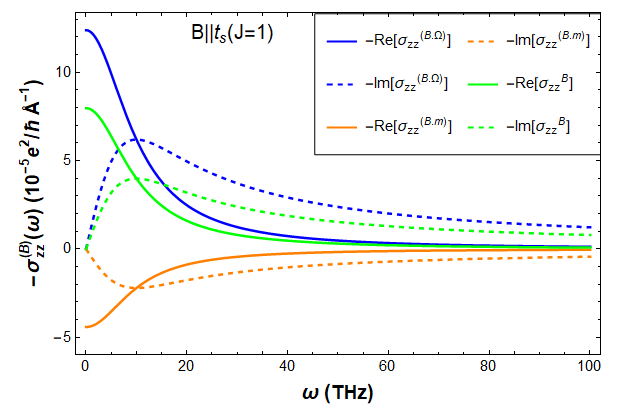} &
\includegraphics[width=0.35\linewidth]{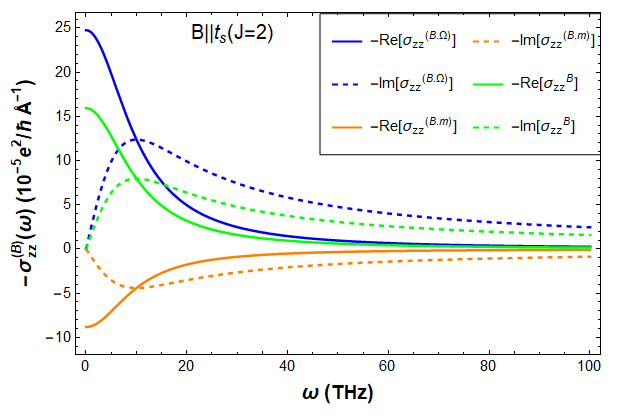} &
\includegraphics[width=0.35\linewidth]{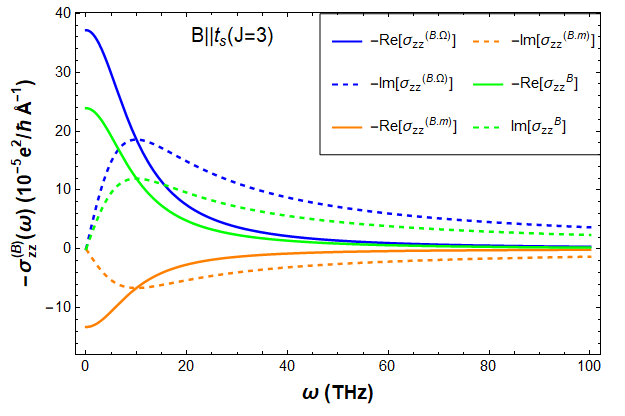} \\
\multicolumn{3}{c}{(b)}\\
\end{tabular}
\caption{ (a)The tilt($t_+$) dependence of the B-linear conductivity $\sigma_{zz}^{(B)}$ for the case of $\textbf{B}\Vert \hat{\bm t}_s$ at B=1 T. (b) The frequency dependence of the B-linear magnetoconductivity at $t_s=0.5$. All the other parameters are identical to those in Fig.(\ref{fig_cond_noB_tilt}).} \
\label{fig_cond_linear_B_zz}\
\end{figure}\
\end{center}\
\twocolumngrid\

\onecolumngrid\
\begin{center}\
\begin{figure}
\begin{tabular}{ccc}
\includegraphics[width=0.35\linewidth]{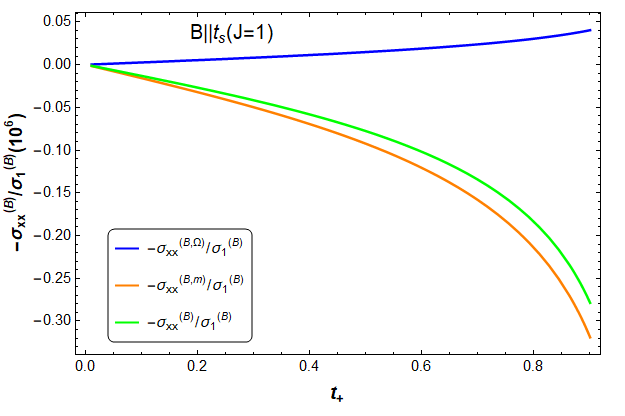} &
\includegraphics[width=0.35\linewidth]{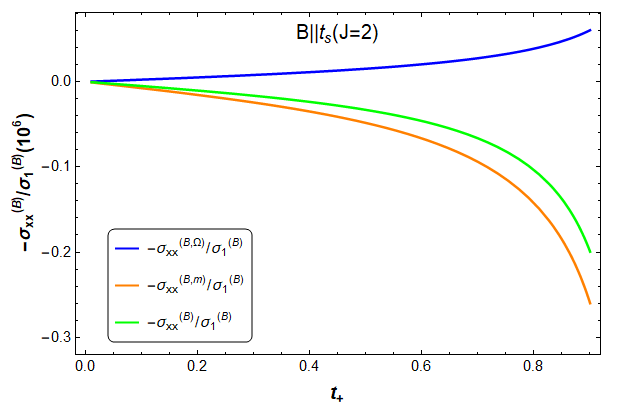} &
\includegraphics[width=0.35\linewidth]{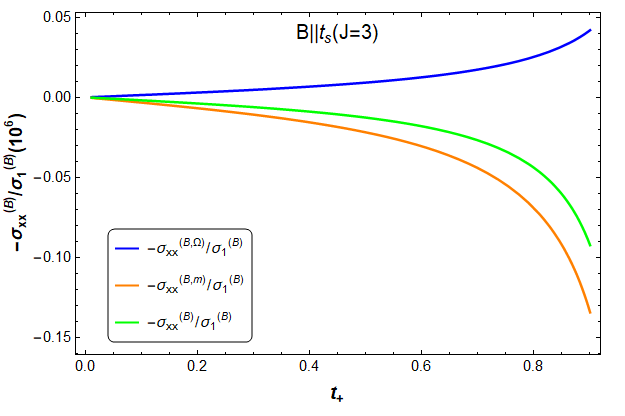} \\ 
\multicolumn{3}{c}{(a)}\\
\includegraphics[width=0.35\linewidth]{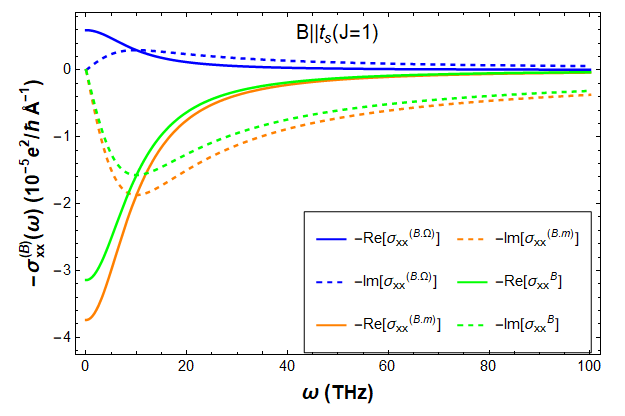} &
\includegraphics[width=0.35\linewidth]{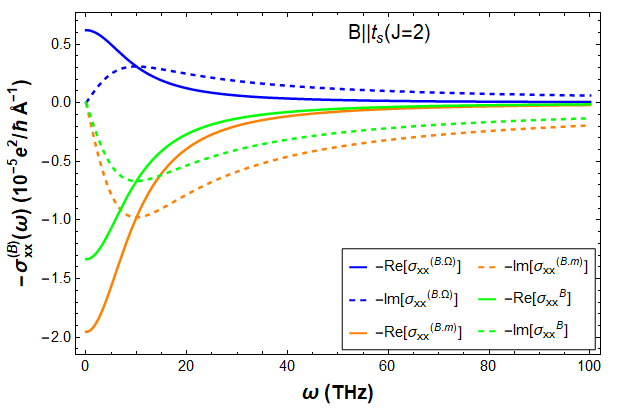} &
\includegraphics[width=0.35\linewidth]{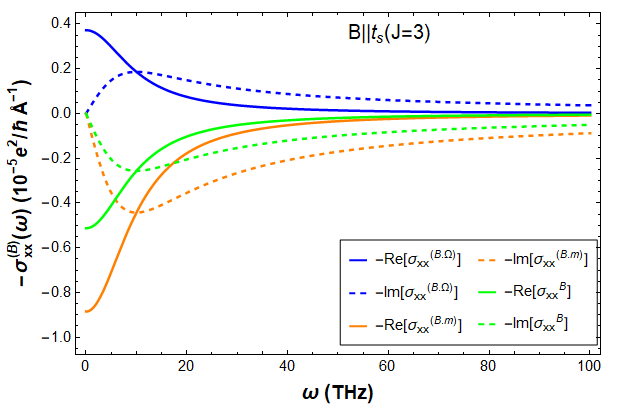} \\
\multicolumn{3}{c}{(b)}\\
\end{tabular}
\caption{ (a) The tilt($t_+$) dependence of the B-linear conductivity $\sigma_{xx}^{(B)}$ for the case of $\textbf{B}\Vert \hat{\bm t}_s$ at B=1 T. (b) The frequency dependence of the B-linear magnetoconductivity at $t_s=0.5$. All the other parameters are identical to those in Fig.(\ref{fig_cond_noB_tilt}).} \
\label{fig_cond_linear_B_xx}\
\end{figure}\
\end{center}\
\twocolumngrid\

Case-I When $\textbf{B}\Vert \hat{\bm t}_s \Vert \hat{\bm z} $\\
For the case of the magnetic field along the tilt direction(B ts ), the off-diagonal magnetoconductivity $\sigma_{ab}^{(B,\Omega)}(a\neq b)$ vanishes since the integrands in Eqs. (\ref{cond_B_Omega}) and (\ref{condcomp}) contain a product of an odd number of the velocity components. The diagonal component $\sigma_{aa}^{(B,\Omega)}$ for small tilt parameter $t_s$ reduces to 
\begin{widetext}
\begin{eqnarray}\label{Berry_B1}
\sigma_{zz}^{(B,\Omega)}&=&\frac{J\mathit{v_{F}} \sigma_1^{(B)}s}{\pi}\Bigl[\frac{(3-5t_s^{2}-3t_s^{4})}{t_s^{3}}+\frac{3(1-2t_s^{2}-4t_s^2)}{2t_s^{4}}\ln\frac{1-t_s}{1+t_s}\Bigr]\\
\sigma_{xx}^{(B,\Omega)}&=&\frac{3J^2\alpha_J^{\frac{2}{J}}\mu^{(2-\frac{2}{J})}\hbar^{(-2+\frac{2}{J})} \sigma_1^{(B)}s}{8\pi^{\frac{1}{2}}\mathit{v_{F}}}\Bigl[(2-3J)t_s\Gamma{(3-1/J){}_2\tilde{F}_1{(2-\frac{1}{J},\frac{5}{2}-\frac{1}{J},\frac{9}{2}-\frac{1}{J},t_s^2)}}\Bigr]
\end{eqnarray}

\noindent where $\sigma_1^{(B)}$=$\frac{e^3\tau B}{(1-i\omega \tau)12\pi\hbar^2}$ 

Using the identity
\begin{eqnarray}
\int_0^{\pi}d\theta \frac{(\sin\theta)^m (\cos\theta)^n}{(1+t_s\cos\theta)^l}&=&
\frac{1}{4} \sqrt{\pi} \, \Gamma\!\left(\frac{1+m}{2}\right) \Bigg[
2 \big(1 + (-1)^n\big) \, \Gamma\!\left(\frac{1+n}{2}\right) \, 
{}_3\tilde{F}_2\left(
\frac{1+l}{2}, \frac{l}{2}, \frac{1+n}{2};
\frac{1}{2}, \frac{2+m+n}{2}
; t^2 \right) \nonumber\\
&+& \big(-1 + (-1)^n\big) \, l \, t \, \Gamma\!\left(1+\frac{n}{2}\right) \, 
{}_3\tilde{F}_2\left(
\frac{1+l}{2}, \frac{2+l}{2}, \frac{2+n}{2} ;
\frac{3}{2}, \frac{3+m+n}{2}; t^2 \right)
\Bigg]\label{regularized_HGF}
\end{eqnarray}
$
\text{if } \Re(m) > -1 \;\&\&\; \Re(n) > -1$
where $_{n_1}\tilde{F}_{n_2}({a_1,a_2,...,a_{n1}};{b_1,b_2,...,b_{n1}};X)$ is the regularized hypergeometric function\cite{weisstein2003regularized}.

Via the similar calculation, Eq.(\ref{condcomp}) becomes\\

\begin{eqnarray}\label{orbital_B}
\sigma_{zz}^{(B,m)}(\omega)&=&\frac{J\mathit{v_F} \sigma_1^{(B)}s}{\pi}\Bigl[\frac{-3+5t_s^2}{t_s^3}-\frac{3(-1+t_s^2)^2}{2t_s^4}\ln{\frac{1-t_s}{1+t_s}}\Bigr]\\
\sigma_{xx}^{(B,m)}(\omega)&=&\frac{J s \alpha_J^{2/J} \mu^{2-\frac{2}{J}} \hbar^{-2+\frac{2}{J}}}
{4 \pi^{\frac{1}{2}} t_s\mathit{ v_F}\Gamma\!\left[\frac{3}{2} - \frac{1}{J}\right]}\Gamma\!\left[2 - \frac{1}{J}\right]\sigma_1^{(B)}
\Bigg[2 J (-1 + 2 J) (1 - t_s^{2})^{-1+\frac{1}{J}}\nonumber\\&+&(t_s^{2})^{-\frac{5}{2}+\frac{1}{J}} \Big(-2 + J (9 - 10 J+ 3 (1 + 2 J) t^{2})\Big)\text{Beta}\!\left[t_s^{2},\, \frac{5}{2} - \frac{1}{J},\, -1 + \frac{1}{J}\right]\Bigg] 
\end{eqnarray}
\end{widetext}

In this case, the above components obey the power laws $\sigma_{zz}^{(B,\Omega)}(\omega), \sigma_{zz}^{(B,m)}(\omega) \propto J\mathit{v_F}\alpha_J^0  $ and $\sigma_{xx}^{(B,\Omega)} (\omega), \sigma_{xx}^{(B,m)}(\omega)\propto \frac{\mu^{2(1-1/J)}\alpha_J^{2/J}}{\mathit{v_F}}$. Figure (\ref{fig_cond_linear_B_zz}a) and (\ref{fig_cond_linear_B_zz}b) illustrate the tilt $t_s$ and frequency dependence of B-linear magnetoconductivity of $\sigma_{zz}^{(B)}$ and  respectively. The orbital magnetic moment has a negative contribution that diminishes with increasing $t_+$, partially cancelling out the Berry curvature's contribution to magnetoconductivity. The total magnetoconductivity tends to slow down at large $t_+$. For $\sigma_{xx}^{(B)}$, the OMM part is dominated over BC part. Thus, the overall conductivity has changed its behavior as reflected in  Fig.(\ref{fig_cond_linear_B_xx}a). Fig.(\ref{fig_cond_linear_B_xx}b) illustrate the variation of  $\sigma_{xx}^{(B)}$ with THz incoming light. \\ 

\onecolumngrid\
\begin{center}\
\begin{figure}
\begin{tabular}{ccc}
\includegraphics[width=0.35\linewidth]{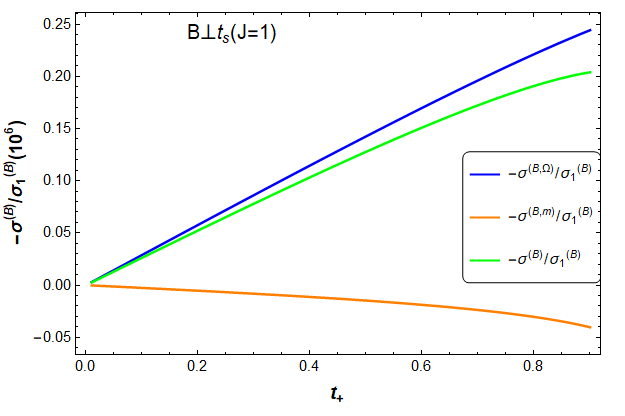} &
\includegraphics[width=0.35\linewidth]{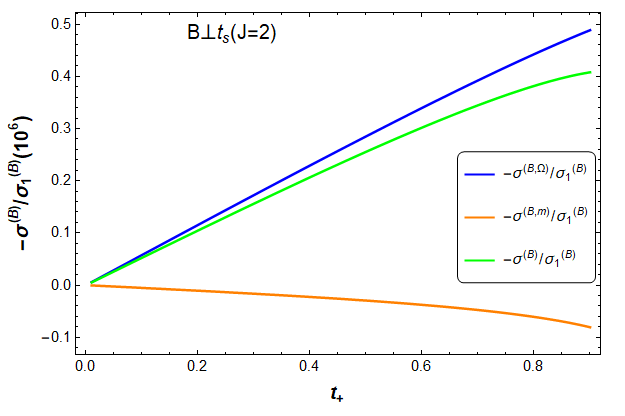} &
\includegraphics[width=0.35\linewidth]{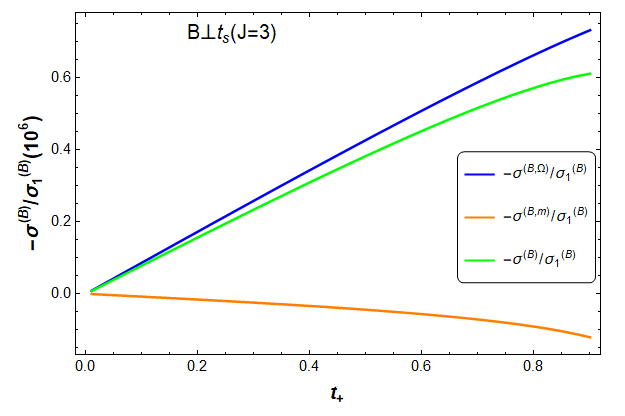} \\ 
\multicolumn{3}{c}{(a)}\\
\includegraphics[width=0.35\linewidth]{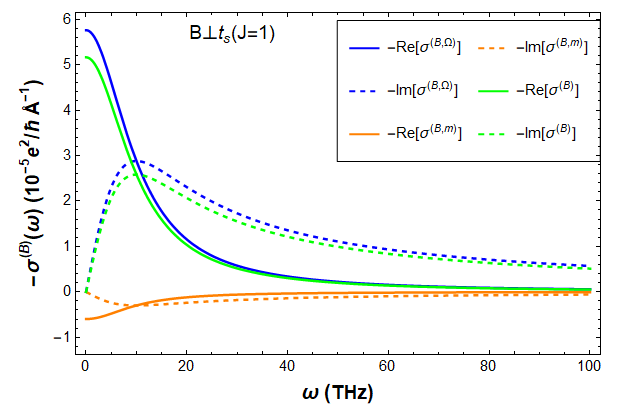} &
\includegraphics[width=0.35\linewidth]{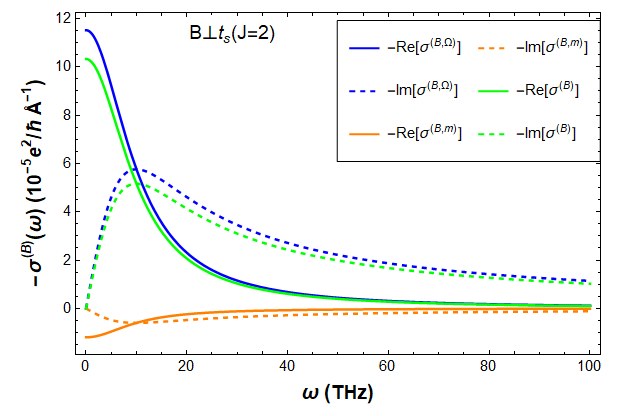} &
\includegraphics[width=0.35\linewidth]{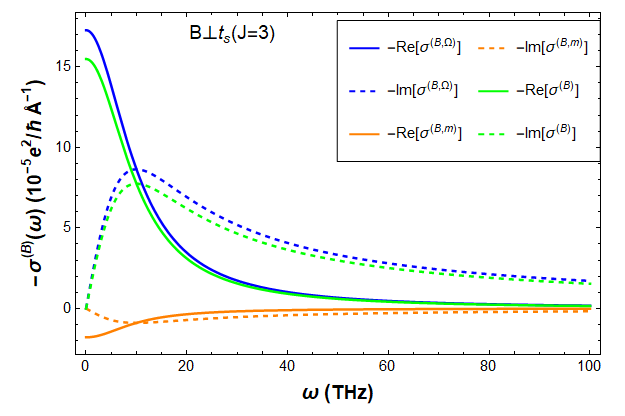} \\
\multicolumn{3}{c}{(b)}\\
\end{tabular}
\caption{ (a) The tilt($t_+$) dependence of the B-linear conductivity $\sigma^{(B)}$ for the case of $\textbf{B}\perp  \hat{\bm t}_s $ at B=1 T. (b) The frequency dependence of the B-linear magnetoconductivity at $t_s=0.5$. All the other parameters are identical to those in Fig.(\ref{fig_cond_noB_tilt}).} \
\label{fig_cond_linear_B_per}\
\end{figure}\
\end{center}\
\twocolumngrid\

Case-II For $\textbf{B}\perp  \hat{\bm t}_s (\Vert \hat{\bm z})$\\
We can represent the magnetic field in the x-y plane as
$\textbf{B}=B(\hat{i}\cos\gamma +\hat{j}\sin \gamma )$ where $\gamma$ is the angle between magnetic field $ \textbf{B} $ and x-axis. In this case, longitudinal components of conductivities are zero and we get the following expressions for planar Hall conductivities \cite{gao2022suppression, das2019linear}
\begin{equation}
\sigma_{xz}^{(B)}(\omega)=[\sigma^{(B,\Omega)}(\omega)+\sigma^{(B,m)}(\omega)]\cos \gamma
\end{equation}

\begin{equation}
\sigma_{yz}^{(B)}(\omega)=[\sigma^{(B,\Omega)}(\omega)+\sigma^{(B,m)}(\omega)]\sin \gamma
\end{equation}

The linear B planar Hall effect due to Berry curvature has been studied in the literature \cite{ghosh2024direction,onofre2024electric}. The OMM contribution is our result.

\begin{widetext}
\begin{eqnarray}
\sigma^{(B,\Omega)}_{xz}(\omega)=\frac{\tau}{\hbar
(1-i\omega \tau)}\frac{e^3}{(2\pi)^3}\int d^3k [\mathit{v}_z^s B \cos \gamma (\mathit{v}_x^s\Omega_{ky}+\mathit{v}_z^s\Omega_{kz})-\mathit{v}_z^s B \sin \gamma \mathit{v}_x^s\Omega_{ky}]\Bigl(-\frac{\partial f_0^s}{\partial \epsilon_{\bm k}^s}\Bigr)
\end{eqnarray}

\begin{eqnarray}
\sigma^{(B,m)}_{xz}(\omega)=\frac{\tau}{\hbar(1-i\omega \tau)}\frac{e^2}{(2\pi)^3}\int d^3k \Bigl[ B \cos \gamma \Bigl(\frac{\partial \mathit{v}_x^s}{\partial k_z}m_{kx}^s-v_z^s\frac{\partial \mathit{m}_{kx}^s}{\partial k_z}\Bigr)- B \sin \gamma \Bigl(\frac{\partial \mathit{v}_x^s}{\partial k_z}m_{ky}^s-v_z^s\frac{\partial \mathit{m}_{ky}^s}{\partial k_x}\Bigr)\Bigr]\Bigl(-\frac{\partial f_0^s}{\partial \epsilon_{\bm k}^s}\Bigr)
\end{eqnarray}
\end{widetext}
The second term of both the above equations contributes to zero. One can obtain

\begin{eqnarray}
\sigma^{(B,\Omega)}&=&\frac{J\mathit{v_F} \sigma_1^{(B)}s}{2\pi}\Bigl[\frac{-3+5t_s^{2}-6t_s^{4}}{t_s^{3}}-\frac{3(1-t_s^2)^{2}}{2t_s^{4}}\ln\frac{1-t_s}{1+t_s}\Bigr]\nonumber\\
\end{eqnarray}
\begin{eqnarray}
\sigma^{(B,m)}&=&\frac{J\mathit{v_F} \sigma_1^{(B)}s}{2\pi}\Bigl[\frac{3-2t_s^{2}}{t_s^{3}}+\frac{3(1-t_s^{2})}{2t_s^{4}}\ln\frac{1-t_s}{1+t_s}\Bigr]\nonumber\\ \label{linearB_cond}
\end{eqnarray}

\noindent Therefore, the Hall conductivity components $\sigma^{(B,\Omega)} $ and  $\sigma^{(B,m)}$ are proportional to topological charge J and independent of Fermi energy $\mu$ but they have different tilt dependence.\\

Eqs. (\ref{Berry_B1})–(\ref{linearB_cond}) show that B-linear magnetoconductivity in tilted multi-Weyl semimetals is an odd function of $t_s$. The Nielsen-Ninomiya theorem \cite{nielsen1981absence,nielsen1981absence} states that Weyl nodes with opposite chirality always form pairs. The system's overall magnetoconductivity is calculated by adding all Weyl nodes together. In the case of $t_{+} =t_{-}$, where the tilt inversion symmetry is broken, the contribution of the Weyl node to magnetoconductivity has the opposite sign for the opposite chirality, yielding $\sigma^{(B)}_{ab}(\omega)$ = 0. Whereas, for the case of  $t_{+} =-t_{-}$, where the tilt inversion symmetry is unbroken, each Weyl node contributes equally to the magnetoconductivity, resulting in nonzero magnetoconductivity. Figure (\ref{fig_cond_linear_B_per}a) illustrates the B-linear magnetoconductivity as a function of tilt $t_+$. The orbital magnetic moment has a negative contribution that diminishes with increasing $t_+$, partially cancelling out the Berry curvature's contribution to magnetoconductivity. The total magnetoconductivity tends to slow down at large $t_+$. Fig.(\ref{fig_cond_linear_B_per}b) illustrates how B-linear magnetoconductivity varies with THz incoming light. We have estimated the values of $\sigma_{ab}^{(B)}$ in Table(\ref{tab:table3}).

\subsection{Calculations of quadratic-B contribution to the conductivity $\sigma_{ab}^{(B^2)}$}

Substituting Eq.(\ref{disf}) with $ \textbf{B} $ terms up to second order into the first term of Eq.(\ref{cur_den}), we get conductivities components

\begin{equation}\label{quadB_cond}
\sigma_{ab}^{(B^2)}(\omega)=\sigma_{ab}^{(B^2,\Omega)}(\omega)+\sigma_{ab}^{(B^2,m)}(\omega)
\end{equation}

where

\begin{widetext}
\begin{equation}
\begin{split}
\sigma_{ab}^{(B^2,\Omega)}(\omega)=& \frac{\tau}{\hbar^2(1-i\omega \tau)}\frac{e^4}{(2\pi)^3}\int d^3k [\mathit{v}_a^s \mathit{v}_b^s(\bm\Omega_{\bm k}^s \cdot \textbf{B})^2-(\mathit{v}_a^s B_b+\mathit{v}_b^s B_a)(\bm\Omega_{\bm k}^s \cdot \bm{\mathit{v}}_{\bm k}^s )(\bm\Omega_{\bm k}^s \cdot \textbf{B})\\& + B_a B_b (\bm\Omega_{\bm k}^s \cdot \bm{\mathit{v}}_{\bm k}^s )^2]\Bigl(-\frac{\partial f_0^s}{\partial \epsilon_{\bm k}^s}\Bigr)\end{split}
\end{equation}

\begin{equation}
\begin{split}
\sigma_{ab}^{(B^2,m)}(\omega)=&\frac{\tau}{\hbar^2(1-i\omega \tau)}\frac{e^3}{(2\pi)^3}\int d^3k\Bigl[\frac{\partial}{\partial \textbf{k}}\cdot[\mathit{v}_a^s B_b \bm\Omega_{\bm k}^s](\textbf{m}_{\bm k}^s\cdot \textbf{B})-\frac{\partial [\mathit{v}_a^s(\bm\Omega_{\bm k}^s \cdot \textbf{B}) ]}{\partial k_b}(\textbf{m}_{\bm k}^s\cdot \textbf{B})+\frac{\partial [B_a^s(\bm\Omega_{\bm k}^s \cdot \bm{\mathit{v}}_{\bm k}^s ) ]}{\partial k_b}(\textbf{m}_{\bm k}^s\cdot \textbf{B})+ \\& \frac{\partial (\textbf{m}_{\bm k}^s\cdot \textbf{B})}{\partial k_a}(\bm\Omega_{\bm k}^s \cdot \textbf{B})\mathit{v}_b^s-\frac{\partial (\textbf{m}_{\bm k}^s\cdot \textbf{B})}{\partial k_a}(\bm\Omega_{\bm k}^s \cdot \bm{\mathit{v}}_{\bm k}^s )B_b^s -\frac{\partial^2 (\textbf{m}_{\bm k}^s\cdot \textbf{B})}{e \partial k_a \partial k_b}(\textbf{m}_{\bm k}^s\cdot \textbf{B})-B_a\frac{\partial (\textbf{m}_{\bm k}^s\cdot \textbf{B})}{\partial \bm k}\cdot \bm\Omega_{\bm k}^s \mathit{v}_b^s \Bigr]\Bigl(-\frac{\partial f_0^s}{\partial \epsilon_{\bm k}^s}\Bigr)
\end{split}
\end{equation}
\end{widetext}

\noindent where $\sigma_{ab}^{(B^2,\Omega)}(\omega)$ and $\sigma_{ab}^{(B^2,m)}(\omega)$
are Berry-curvature $\bm{\Omega}_{\bm k}^s$ and the orbital magnetic moment $\textbf{m}_{\bm k}^s$ dependent conductivities, respectively. In order to decode the information on $\bm{\Omega}_{\bm k}^s$ and ${\bm m}^s_{\bm k}$ in Eq.(\ref{quadB_cond}), we consider the two case in the following:\\

Case-I When $\textbf{B}\Vert \hat{\bm t}_s \Vert \hat{\bm z} $\\
 In this case, Hall conductivities components are zero and one can get the following expressions for longitudinal components

\begin{widetext}
\begin{eqnarray}
\sigma_{zz}^{(B^2,\Omega)}(\omega)=\frac{J^3\alpha_J^{\frac{2}{J}}\mathit{v}_F\mu^{\frac{-2}{J}}\hbar^{-1+\frac{2}{J}}\sigma_1^{(B^2)}}{4\pi^{\frac{1}{2}}}\Bigr[\Gamma(4-1/J){}_2\tilde{F}_1{(\frac{1}{2}-\frac{1}{J},\frac{-1+J}{J},\frac{9}{2}-\frac{1}{J},t_s^2)}\Bigr]
\end{eqnarray}

\begin{equation}
\begin{split}    
\sigma_{xx}^{(B^2,\Omega)}(\omega)=\frac{J^5\alpha_J^{\frac{4}{J}}\mu^{2-\frac{4}{J}}\hbar^{-3+\frac{4}{J}}\sigma_1^{(B^2)}}{16\mathit{v}_F}\bigg[\Gamma{\Big(4-\frac{2}{J}\Big)}{}_3\tilde{F}_2\bigg\{{\Big(\frac{3}{2},\frac{3}{2}-\frac{2}{J},2-\frac{2}{J}\Big)},{\Big(\frac{1}{2},\frac{11}{2},\frac{-2}{J}\Big),t_s^2}\bigg\}\bigg] 
\end{split}   
\end{equation} 
\end{widetext}

\noindent where $\sigma_1^{(B^2)}=\frac{e^4 \tau  B^2 }{4\pi\hbar^{2}(1-i\omega \tau)}$. Therefore, the above components follow the power laws $\sigma_{zz}^{(B^2,\Omega)}(\omega) \propto \frac{J^3\alpha_J^{\frac{2}{J}}\mathit{v}_F}{\mu^{\frac{2}{J}}}$, $\sigma_{xx}^{(B^2,\Omega)}(\omega) \propto \frac{J^5\alpha_J^{\frac{4}{J}}}{\mu^2(\frac{2}{J}-1)\mathit{v}_F}$.\\

Considering the effect of the orbital magnetic moment, we have

\begin{widetext}
\begin{eqnarray}
\sigma_{zz}^{(B^2,m)}(\omega)&=&\frac{\alpha_J^{\frac{2}{J}} \mathit{v}_F\mu^{-\frac{2}{J}}\hbar^{-1+\frac{2}{J}}\sigma_1^{(B^2)}}{16\pi^{\frac{1}{2}}}\frac{\Gamma(2-\frac{1}{J})}{\Gamma(\frac{9}{2}-\frac{1}{J})}\bigg[(-2+7J)\Big\{(2-5J)J{}_2F_1{\Big(\frac{1}{2}-\frac{1}{J},\frac{-1+J}{J},\frac{5}{2}-\frac{1}{J},t_s^2\Big)}\nonumber\\
&+&(-2+J)^2t_s^2{}_2F_1{\Big(\frac{3}{2}-\frac{1}{J},\frac{-1+J}{J},\frac{7}{2}-\frac{1}{J},t_s^2\Big)}\Big\}+2J(1+2J)(-2+7J){}_3F_2\bigg\{{\Big(\frac{3}{2},\frac{1}{2}-\frac{1}{J},1-\frac{1}{J}\Big)},{\Big(\frac{1}{2},\frac{7}{2}-\frac{1}{J}\Big),t_s^2}\bigg\}\nonumber\\
&-&18J^3{}_3F_2\bigg\{{\Big(\frac{5}{2},\frac{1}{2}-\frac{1}{J},1-\frac{1}{J}\Big)},{\Big(\frac{1}{2},\frac{9}{2}-\frac{1}{J}\Big),t_s^2}\bigg\}\bigg]
\end{eqnarray}

A closed-form expression for $\sigma_{xx}^{(B^2,m)}(\omega)$ for arbitrary $J$ cannot be obtained because the condition $\Re(n)>-1$ in Eq.~(\ref{regularized_HGF}) is satisfied only for $J>2$. We therefore give the exact expressions separately for each monopole charge.

\begin{align}
\sigma_{xx}^{(B^2,m)}(\omega)=&-\frac{\mathit{v_F}^3\hbar}{30\pi\mu^2},(J=1)\\
\sigma_{xx}^{(B^2,m)}(\omega)=&\frac{8\alpha_2^2}{\pi\mathit{v_F\hbar}}
\Big[\frac{(30+20t_s^2+t_s^4)}{5t_s^6}+\frac{(6-6t_s^2+t_s^4)}{2t_s^7}\ln\left(\frac{1-t_s}{1+t_s}\right)\Big],(J=2)\\
\sigma_{xx}^{(B^2,m)}(\omega)=&-\frac{2187 \, \alpha_3^{4/3} \, \mu^{2/3} \, \Gamma\!\left(\frac{10}{3}\right)}
{1204280 \pi^{\frac{1}{2}}\, \mathit{v_F} \, \hbar^{5/3} \, \Gamma\!\left(\frac{5}{6}\right)}
\Bigg[782 \, {}_3F_2\left\{\left(\frac{5}{6},\, \frac{4}{3},\, \frac{3}{2}\right), \left(\frac{1}{2}, \frac{17}{6}\right), t_s^2 \right\}\nonumber\\
&- 4347 \, {}_3F_2\!\left\{\left( \frac{5}{6},\, \frac{4}{3},\, \frac{5}{2}\right)\left(\frac{1}{2},\, \frac{23}{6}\right),t_s^2 \right\}+ 3645 \, {}_3F_2\!\left\{\left(\frac{5}{6},\, \frac{4}{3},\, \frac{7}{2}\right),\left(\frac{1}{2},\, \frac{29}{6}\right), t_s^2 \right\}
\Bigg],(J=3)
\end{align}
\end{widetext}

Therefore,the above components follow the power laws $\sigma_{zz}^{(B^2,m)}(\omega) \propto \frac{\alpha_J^{\frac{2}{J}}\mathit{v}_F}{\mu^{\frac{2}{J}}}$ and $\sigma_{xx}^{(B^2,m)} (\omega)\propto \frac{\alpha_J^{\frac{4}{J}}\mu^{2(1-\frac{2}{J})}}{\mathit{v}_F} $. In the case of single WSM, the total conductivity component $\sigma_{zz}^{(B^2)}(\omega)$ is suppressed(enhanced) below(above) the value of $t_0$ as shown in Fig.(\ref{fig_cond_quadratic_B_zz}a)(see ref.\cite{gao2022suppression}). For double and triple WSMs, the total conductivity components $\sigma_{zz}^{(B^2)}(\omega)$ are suppressed almost constantly but largely with tilt parameter $t_s$ as compared with the J=1 case, as illustrated in Fig.(\ref{fig_cond_quadratic_B_zz}a).  Further, in the case of single-WSM,
$\sigma_{xx}^{(B^2,\Omega)}(\omega)$ and $\sigma_{xx}^{(B^2,m)}(\omega)$ have equal and opposite values, thus the total conductivity $\sigma_{xx}^{(B^2)}(\omega)$ is equal to zero  \cite{gao2022suppression}. There are no such cancellations for higher J semimetals. For double WSMs, the total conductivity is enhanced(suppressed) below(above) the critical value $t_0$ of the tilt parameter and dominated by BC(OMM) as shown in Fig.(\ref{fig_cond_quadratic_B_zz}b). In case of triple WSMs, the total conductivity is suppressed and dominated by the orbital contribution(see Fig.(\ref{fig_cond_quadratic_B_zz}b).\\

Case-II  For $\textbf{B}\perp  \hat{\bm t}_s (\Vert \hat{\bm z})$\\
In this case, both diagonal as well as planar Hall conductivities are non-zero. The longitudinal components have the following expressions

\begin{widetext}
\begin{equation}
\begin{split}
\sigma_{xx}^{(B^2,\Omega)}(\omega)= &\frac{\mathit{v_F}\, \alpha_J^{2/J} \mu^{-2/J} \hbar^{-1 + \frac{2}{J}}\sigma_1^{(B^2)}}{960\,J\, \pi^{1/2}} \Bigg[ \frac{1}{t^4} \cos^2\gamma \; \Gamma\!\left(2 - \frac{1}{J}\right)
\Bigg( \big(4 + J(-32 + 89J - 97J^2 + 30J^3) - 4(-2+J)(-1+2J)\\&(-7+9J)t_s^2+ J(207 + J(-115 + 122J))t_s^4 + 120J^2 t_s^6 \big)
\,{}_2\tilde{F}_1\!\left(\frac{1}{2} - \frac{1}{J}, \frac{-1+J}{J}; \frac{5}{2} - \frac{1}{J}; t_s^2\right)- (-2+J)(-1+t_s^2)\\&\big(2 + J(-15 + 37J - 30J^2 + (25 + J(-77 + 54J))t_s^2 + 8J(9+2J)t_s^4)\big) {}_2\tilde{F}_1\!\left(\frac{3}{2} - \frac{1}{J}, \frac{-1+J}{J}; \frac{5}{2} - \frac{1}{J}; t_s^2\right) \Bigg) \\
& + 30 J^4 \Gamma\!\left(4 - \frac{1}{J}\right)
{}_2\tilde{F}_1\!\left(\frac{1}{2} - \frac{1}{J}, \frac{-1+J}{J}; \frac{9}{2} - \frac{1}{J}; t_s^2\right) \sin^2\gamma \Bigg]
\end{split}
\end{equation}
\begin{eqnarray}
\sigma_{zz}^{(B^2,\Omega)}(\omega)=\frac{J\mathit{v_F^3\hbar}\sigma_1^{(B^2)}}{30\pi\mu^2}(1+7t_s^2)
\end{eqnarray}

The off-diagonal components of magnetoconductivities
\begin{equation}
\begin{split}
\sigma_{xy}^{(B^2,\Omega)}(\omega) = &\frac{\mathit{v_F}\, \alpha_J^{2/J} \mu^{-2/J} \hbar^{-1 + \frac{2}{J}}\sigma_1^{(B^2)}}{1440\, J\, \pi^{1/2}\, t_s^4} \Gamma\!\left(2 - \frac{1}{J}\right) \Bigg[ \Big(4 + J \big(8(-4 + 9 t_s^2) + J\big(89 - 97J + 30J^2- 36(7 + J(-7 + 2J))t_s^2\\& + 3(101 + J(-45 + 46J))t_s^4 + 180J t_s^6\big)\big)\Big){}_2\tilde{F}_1\!\left(\frac{1}{2} - \frac{1}{J}, \frac{-1+J}{J}; \frac{5}{2} - \frac{1}{J}; t_s^2\right)- (-2+J)(-1+t^2)\big(2 + J\big(-15 \\&+ 33t_s^2 + J\big(37 - 30J + 3(-31 + 18J)t_s^2 + 12(9+2J)t_s^4\big)\big)\big){}_2\tilde{F}_1\!\left(\frac{3}{2} - \frac{1}{J}, \frac{-1+J}{J}; \frac{5}{2} - \frac{1}{J}; t_s^2\right) \Bigg] \cos\gamma\, \sin\gamma
\end{split}
\end{equation}
\end{widetext}

\onecolumngrid\
\begin{center}\
\begin{figure}
\begin{tabular}{ccc}
\includegraphics[width=0.36\linewidth]{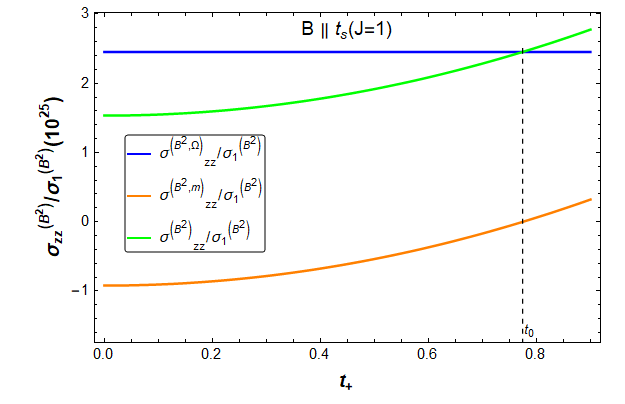} &
\includegraphics[width=0.36\linewidth]{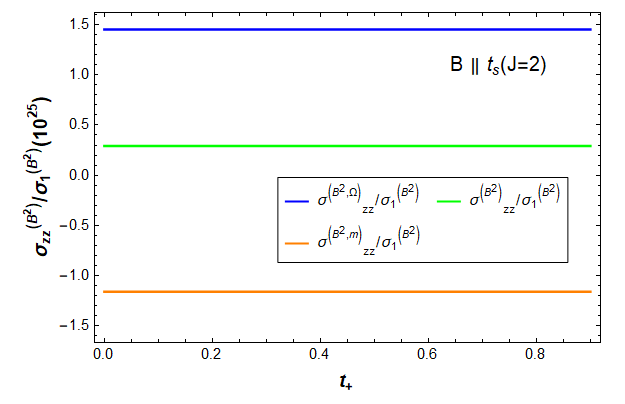} &
\includegraphics[width=0.36\linewidth]{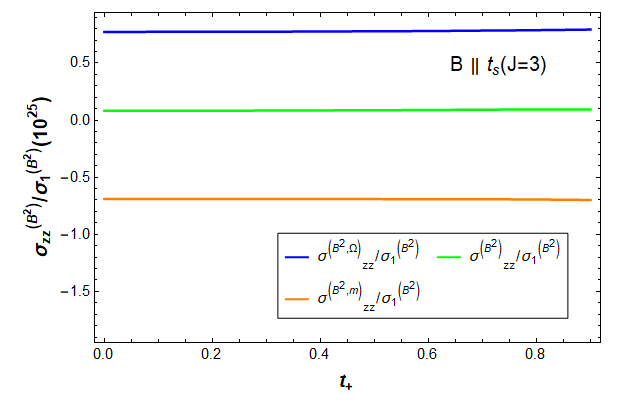} \\ 
\multicolumn{3}{c}{(a)}\\
\includegraphics[width=0.36\linewidth]{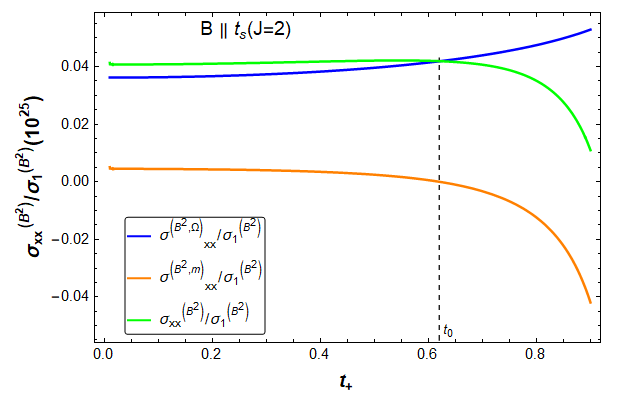} &
\includegraphics[width=0.36\linewidth]{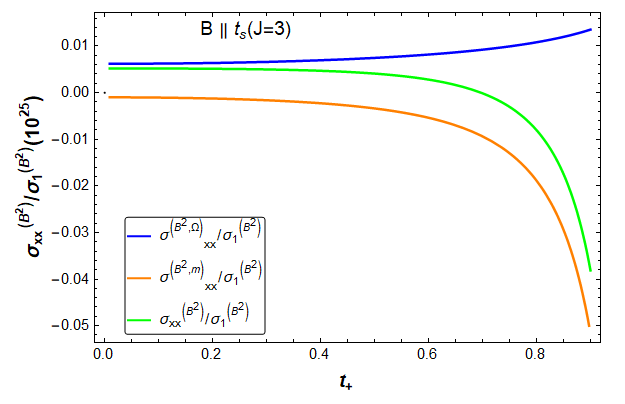} \\
\multicolumn{2}{c}{(b)}\\
\end{tabular}
\caption{ (a) The tilt $t_+$ dependence of the quadratic B longitudinal conductivity $\sigma_{zz}^{(B^2)}$. (b) The tilt $t_+$ dependence of the quadratic B longitudinal conductivity  $\sigma_{xx}^{(B^2)}$  for the case of  $\bm B \parallel \hat{\bm t}_s$ at B=1 T. All the other parameters are identical to those in Fig.(\ref{fig_cond_noB_tilt}).} \
\label{fig_cond_quadratic_B_zz}\
\end{figure}\
\end{center}\
\twocolumngrid\

The orbital magnetic moment contributions of conductivities

\begin{widetext}
\begin{equation}
\begin{split}
\sigma_{xx}^{(B^2,m)}(\omega) = &\frac{ \mathit{v_F}\, \alpha_J^{2/J} \mu^{-2/J} \hbar^{-1 + \frac{2}{J}}\sigma_1^{(B^2)}}{960\, J\, \pi^{1/2}} \Gamma\!\left(2 - \frac{1}{J}\right) \Bigg[ \frac{2}{t_s^4} \cos^2\gamma \Bigg( \big(-12 + J(96 - 3J(89 + J(-97 + 30J)))\\&+ (-2+J)(-1+2J)(-7+13J)t_s^2 + 2J(-11 + J(-75 + 44J))t_s^4 \big)  {}_2\tilde{F}_1\!\left(\frac{1}{2} - \frac{1}{J}, \frac{-1+J}{J}; \frac{5}{2} - \frac{1}{J}; t_s^2\right)\\& + (-2+J)(-1+t_s)(1+t_s)\big(6 + J\big(-45 + 3(37 - 30J)J - 2(1+J(-9+14J))t^2 + 16J(2+J)t_s^4\big)\big)\\& {}_2\tilde{F}_1\!\left(\frac{3}{2} - \frac{1}{J}, \frac{-1+J}{J}; \frac{5}{2} - \frac{1}{J}; t_s^2\right) \Bigg) + 15J^3 \Bigg( -4\, {}_2\tilde{F}_1\!\left(\frac{1}{2} - \frac{1}{J}, \frac{-1+J}{J}; \frac{5}{2} - \frac{1}{J}; t_s^2\right) + \sqrt{\pi} \bigg( 2(3+J)\, \\&{}_3\tilde{F}_2\!\left(\frac{3}{2}, \frac{1}{2} - \frac{1}{J}, \frac{-1+J}{J}; \frac{1}{2}, \frac{7}{2} - \frac{1}{J}; t_s^2\right)- 9J\, {}_3\tilde{F}_2\!\left(\frac{5}{2}, \frac{1}{2} - \frac{1}{J}, \frac{-1+J}{J}; \frac{1}{2}, \frac{9}{2} - \frac{1}{J}; t_s^2\right) \bigg) \Bigg) \sin^2\gamma \Bigg]
\end{split}
\end{equation}

\begin{eqnarray}
\sigma_{zz}^{(B^2,m)}(\omega)=\frac{J\mathit{v_F^3\hbar}\sigma_1^{(B^2)}}{30\pi\mu^2}(-1+t_s^2)
\end{eqnarray}

\begin{equation}
\begin{split}
\sigma_{xy}^{(B^2,m)}(\omega) = &\frac{\mathit{v_F}\alpha_J^{2/J} \mu^{-2/J} \hbar^{-1 + \frac{2}{J}}\sigma_1^{(B^2)}}{240\, J\, \pi^{1/2}\, t_s^4} \Gamma\!\left(2 - \frac{1}{J}\right)\Bigg[ \Big(-4 + J\big(32 - 89J + 97J^2 - 30J^3 + (-5+J)(-2+J)(-1+2J)t_s^2\\&+ 2J(-13 - 30J + 22J^2)t_s^4\big)\Big) {}_2\tilde{F}_1\!\left(\frac{1}{2} - \frac{1}{J}, \frac{-1+J}{J}; \frac{5}{2} - \frac{1}{J}; t_s^2\right)+ (-2+J)(-1+t_s^2)\big(2 + J\big(-15 + 2t_s^2\\& + J\big(37 - 30J + 4(1-4J)t_s^2 + 8(2+J)t_s^4\big)\big)\big)\times {}_2\tilde{F}_1\!\left(\frac{3}{2} - \frac{1}{J}, \frac{-1+J}{J}; \frac{5}{2} - \frac{1}{J}; t_s^2\right) \Bigg] \sin\gamma\, \cos\gamma
\end{split}
\end{equation}
\end{widetext}

\onecolumngrid\
\begin{center}\
\begin{figure}
\begin{tabular}{ccc}
\includegraphics[width=0.35\linewidth]{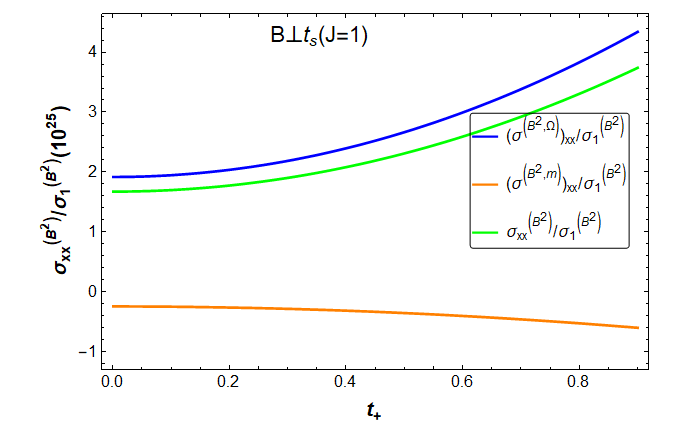} &
\includegraphics[width=0.35\linewidth]{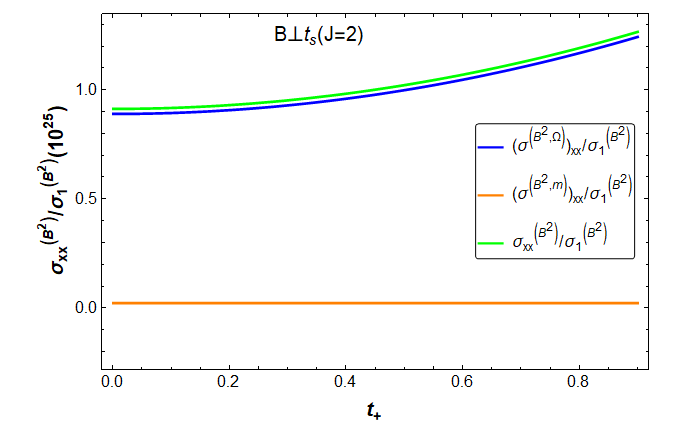} &
\includegraphics[width=0.35\linewidth]{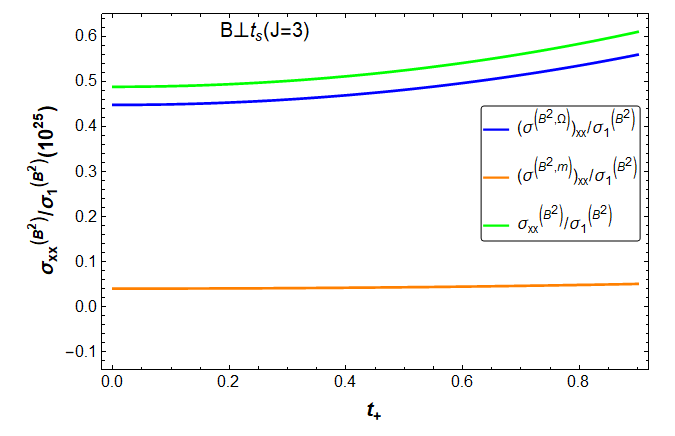} \\ 
\multicolumn{3}{c}{(a)}\\
\includegraphics[width=0.35\linewidth]{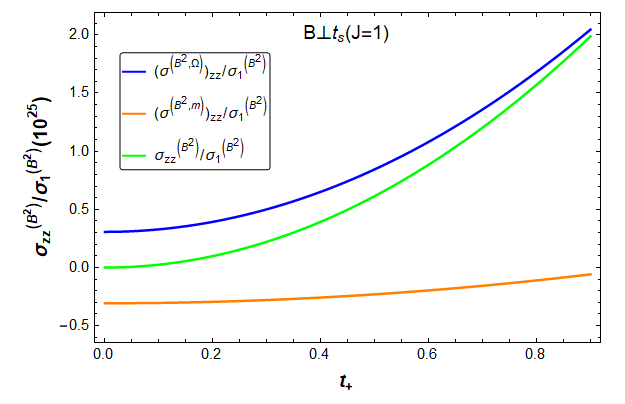} &
\includegraphics[width=0.35\linewidth]{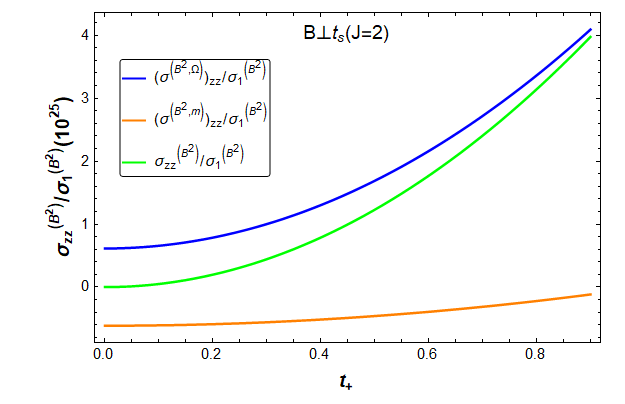} &
\includegraphics[width=0.35\linewidth]{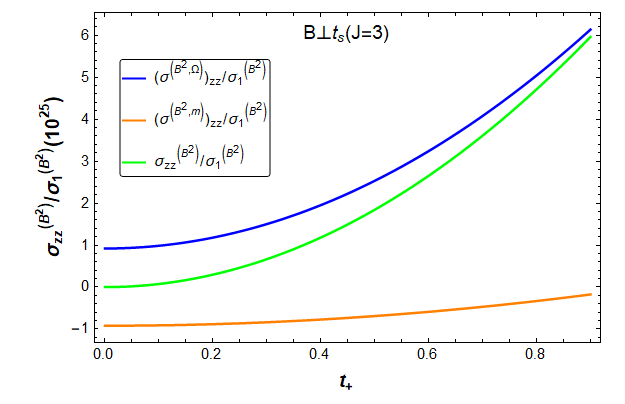} \\
\multicolumn{3}{c}{(b)}\\
\end{tabular}
\caption{ (a) The tilt $t_+$ dependence of the quadratic B conductivity $\sigma_{xx}^{(B^2)}$ at  $\gamma=\pi/6$. (b) The tilt $t_+$ dependence of the quadratic B conductivity $\sigma_{zz}^{(B^2)}$ for the case of  $\bm B \perp \hat{\bm t}_s$ at B=1 T and $\gamma=\pi/4$ . All the other parameters are identical to those in Fig.(\ref{fig_cond_noB_tilt}).} \
\label{fig_cond_quadratic_B_xx_per}\
\end{figure}\
\end{center}\
\twocolumngrid\

\onecolumngrid\
\begin{center}\
\begin{figure}
\begin{tabular}{ccc}
\includegraphics[width=0.35\linewidth]{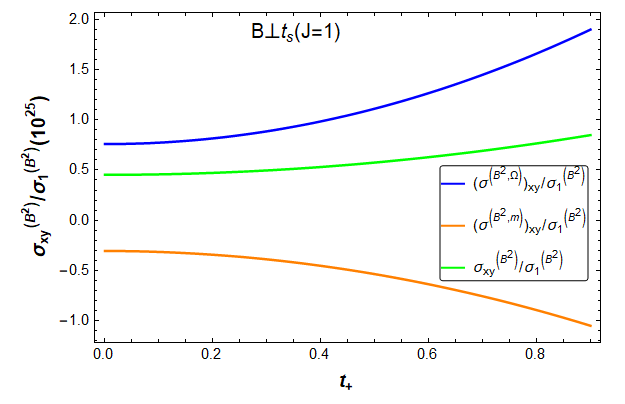} &
\includegraphics[width=0.35\linewidth]{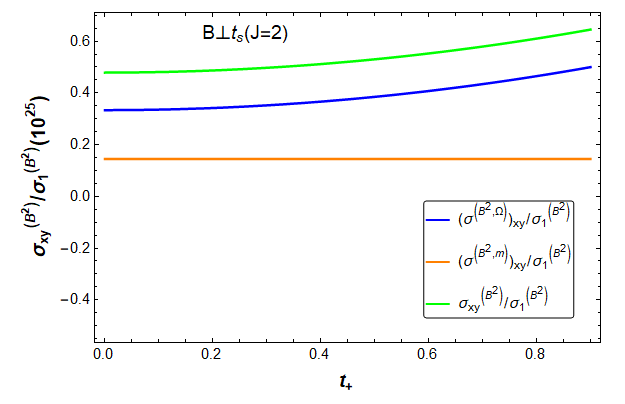} &
\includegraphics[width=0.35\linewidth]{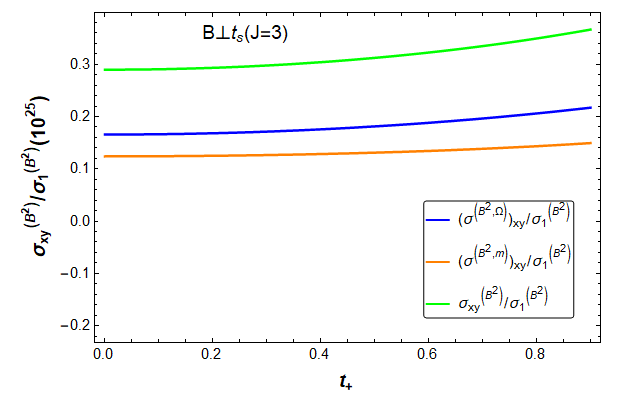} \\ 
\multicolumn{3}{c}{(a)}\\
\includegraphics[width=0.35\linewidth]{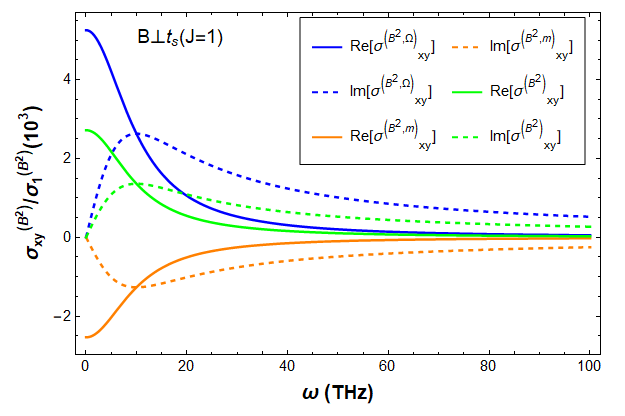} &
\includegraphics[width=0.35\linewidth]{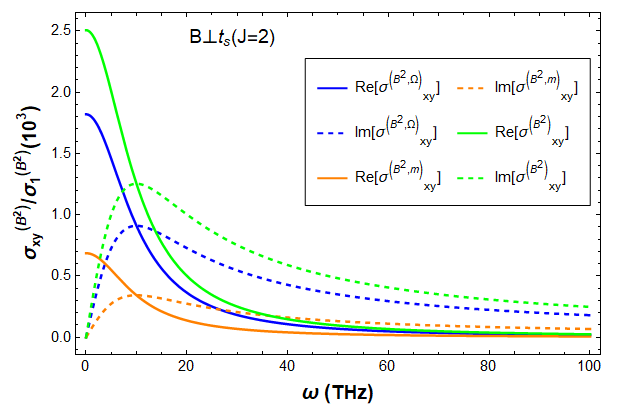} &
\includegraphics[width=0.35\linewidth]{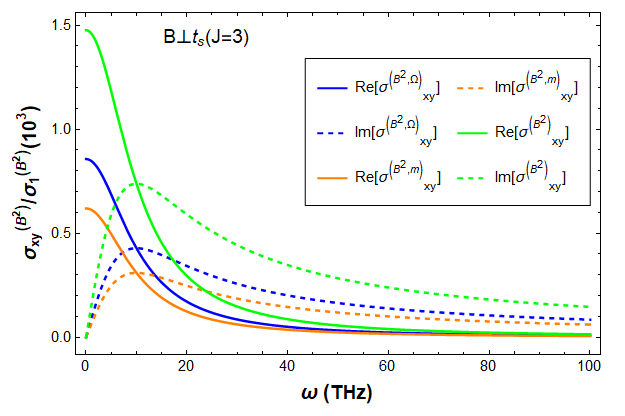} \\
\multicolumn{3}{c}{(b)}\\
\end{tabular}\caption{ (a) The tilt $t_+$ (b) The frequency dependence of the quadratic B planar Hall conductivity $\sigma_{xy}^{(B^2)}$ at $t_s=0.5$ for the case of  $\bm B \perp \hat{\bm t}_s$ at B=1 T and $\gamma=\pi/4$. All the other parameters are identical to those in Fig.(\ref{fig_cond_noB_tilt}).} \
\label{fig_cond_quadratic_B_xy}\
\end{figure}\
\end{center}\
\twocolumngrid\

It is noted that all the other magnetoconductivity components are zero. The preceding conductivity equations are chirality-independent, meaning that Weyl cones with opposite chiralities contribute equally to conductivity. The OMM components also follow the same power laws $\sigma_{zz}^{(B^2,\Omega)}(\omega)$, $\sigma_{zz}^{(B^2,m)}(\omega) \propto \frac{J \mathit{v_F}^3\alpha_J^0}{\mu^2}$ and $\sigma_{xx}^{(B^2,\Omega)} (\omega)$, $\sigma_{xx}^{(B^2,m)} (\omega)$,  $\sigma_{xy}^{(B^2,\Omega)}(\omega)$,$\sigma_{xy}^{(B^2,m)}(\omega) \propto \frac{\alpha_J^{\frac{2}{J}}\mathit{v_F}}{J\mu^{\frac{2}{J}}}$. In this set up, the overall diagonal conductivity is suppressed compared to its Berry curvature parts except the components  $\sigma_{xx}^{(B^2)} (\omega)$  J=2,3[see Fig.(\ref{fig_cond_quadratic_B_xx_per})a]. The planar Hall resistivity in multi-WSMs is estimated in Table(\ref{tab:table3})\\

In the present system, the planar Hall effect can take place \cite{burkov2017giant, nandy2017chiral, ma2019planar, das2019linear, kumar2018planar, yang2019current} and show up as a nonzero conductivity $\sigma_{xy}^{(B^2 )} (\omega)$.  Similar to the diagonal component of conductivity, $\sigma_{xy}^{(B^2 )}(\omega)$  includes contributions from Berry curvature and orbital magnetic moment. The impact of the tilt on the planar Hall magnetoconductivity is depicted in Figure (\ref{fig_cond_quadratic_B_xy}a). The presence of the orbital magnetic moment suppresses the overall planar Hall conductivity in the case of a single WSM. However, for double and triple WSMs, the tilt parameter raises the overall conductivities, respectively, due to the positive contribution of the  orbital magnetic moment. The planar Hall conductivity becomes weaker with topological charges. The planar Hall magnetoconductivity as a function of THz incoming light at  $\gamma= \pi/4$ is displayed in Fig(\ref{fig_cond_quadratic_B_xy}b). The negative contribution of the real and imaginary parts of the orbital  magnetic moment conductivities suppresses the real and imaginary parts of total conductivities in the case of a single WSM. The positive contributions of the real and imaginary parts of orbital magnetic moment conductivities boost the real and imaginary parts of conductivities for double and triple WSMs.\\

Experimental observations of the magnetoconductivity's $B^2$ dependency have been made in materials like GdPtBi and TaP \cite{kumar2018planar, yang2019current}. The band dispersions of TaS and GdPtBi are isotropic and quadratic, respectively. They are therefore excellent choices for single and double WSMs to confirm our findings. Additionally, it has recently been demonstrated that the Berry curvature and chiral anomaly contributions in the material GdPtBi are responsible for a very large planar Hall effect \cite{kumar2018planar}.\\

\begin{figure}
\begin{tabular}{c}
\includegraphics[width=1\linewidth]{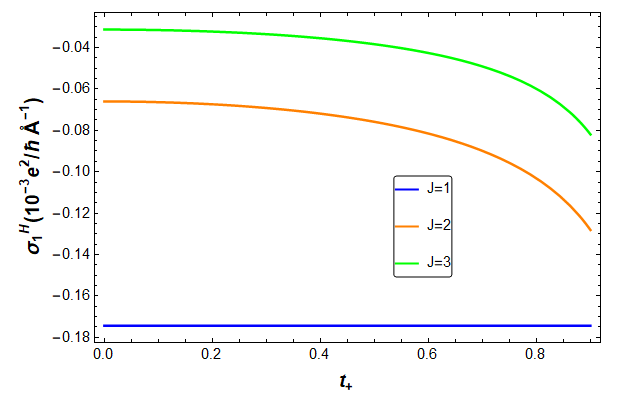} \
\end{tabular}
\caption{The tilt $t_+$  dependence of ordinary Hall conductivity at B=1 T. All the other parameters are identical to those in Fig.(\ref{fig_cond_noB_tilt}).} 
\label{cond_Hall}\
\end{figure}\

\subsection{Hall conductivities $\sigma_{ab}^{(H,0)}$ and $\sigma_{ab}^{(H,B)}$}
The Berry curvature is responsible for the intrinsic Hall effect, as seen in the second term of Eq.(\ref{cur_den}). These calculations yield

\begin{equation}
\sigma_{ab}^{(H)}=\sigma_{ab}^{(H,0)}+\sigma_{ab}^{(H,B)}\label{hall_cond}
\end{equation}

where
\begin{eqnarray}
\sigma_{ab}^{(H,0)}(\omega)&=&-\frac{e^2}{\hbar(2\pi)^3}\epsilon_{abc}\int d^3k \bm\Omega_c^s f_0^s\\
\sigma_{ab}^{(H,B)}(\omega)&=&\frac{e^2}{\hbar(2\pi)^3}\epsilon_{abc}\int d^3k \bm\Omega_c^s(\textbf{m}_{\bm k}^s\cdot \textbf{B})\Bigl(\frac{\partial f_0^s}{\partial \epsilon_{\bm k}^s}\Bigr)
\label{anomalou_hall}
\end{eqnarray}

\noindent where $ \epsilon_{abc} $ is the Levi-Civita symbol with $a, b, c,\in  {x, y, z}$.
In systems with broken time-reversal symmetry, the first term  $ \sigma_{ab}^{(H,0)}(\omega) $ in Eq. (\ref{hall_cond}) does not equal zero \cite{xiao2010berry, gao2022suppression}. This term represents the anomalous Hall effect. The second element,$\sigma_{ab}^{(H,B)}(\omega)$, represents the ordinary Hall conductivity linear in B, which is purely quantum mechanical in nature and cannot be derived from classical Drude theory \cite{gao2014field, das2021intrinsic}. Furthermore, in noncentrosymmetric materials with time-reversal symmetry (TRS), the OMM-induced Hall conductivity, specified by Eq. (\ref{anomalou_hall}), is the only intrinsic Hall response as the anomalous Hall conductivity (AHC) vanishes identically.\\

To illustrate the angular dependence, the magnetic field is written in spherical coordinates,  $ \textbf{B}=(B_x,B_y,B_z)=(B \sin\theta \cos\phi, B \sin\theta \sin\phi, B\cos\theta ) $ with $\theta=\cos^{-1}(B_z/B)$,  $\phi=\tan^{-1}(B_y/B_x)$. For a single multi-Weyl node, the B-linear contribution to the Hall conductivity can be written as 

\begin{equation}
\sigma^{(H,B)}\\
=
\left(
\begin{array}{ccc}
0 &\sigma_1^H \cos \theta &-\sigma_1^H \sin \theta \sin \phi \\
-\sigma_1^H \cos \theta  & 0& \sigma_1^H \sin \theta \cos \phi \\
\sigma_1^H \sin \theta \sin \phi &-\sigma_1^H \sin \theta \cos \phi&0
\end{array}
\right)   \label{eq1}
\end{equation}%

where

\begin{widetext}
\begin{equation}
\sigma_{1}^{H} = - \frac{B\, e^3 J^3 \alpha_J^{2/J} \mu^{\frac{-2+J}{J}} \hbar^{-3 + \frac{2}{J}} }{32\, \pi\, \mathit{v_F}} \Gamma\!\left(2 - \frac{1}{J}\right) {}_3\tilde{F}_2\!\left(\frac{3}{2}, \frac{-1+J}{J}, \frac{3}{2} - \frac{1}{J}; \frac{1}{2}, \frac{7}{2} - \frac{1}{J}; t_s^2\right)
\end{equation}
\end{widetext}

Therefore, the $\sigma_{1}^{H}$ follows the power laws $\sigma_1^H\propto \frac{\alpha_J^{\frac{2}{J}}\mu^{(1-\frac{2}{J})}}{\mathit{v_F}}$. The Hall conductivity is determined by summing over all the Weyl nodes and is independent of the chirality of the valley. The B-linear contribution to the Hall conductivity of larger topological charges $J>1$ depends on the tilt parameter $t_s$, unlike the J=1 WSM (see Fig. (\ref{cond_Hall})).\\

It is important to emphasize that the present treatment differs fundamentally from quantum mechanical descriptions based on Landau-level quantization. The expressions derived above are valid in the weak magnetic fields and small cyclotron frequency $\omega_c$, where the Landau quantization can be ignored. It is valid in the regime  $\hbar\omega_c\ll\mu$($\omega_c$ is cyclotron frequency of multi-WSM) and many Landau levels remain occupied. Under these conditions, the electronic motion is well described by semiclassical wave packets and the Hall response originates primarily from BC corrections to the carrier dynamics. At sufficiently strong magnetic fields, however, discrete Landau levels dominate the electronic spectrum, leading to quantum oscillations and quantum Hall phenomena that cannot be captured within the present framework. Therefore, the semiclassical theory
should be viewed as complementary to Landau-level approaches: it accurately describes
finite-frequency magneto-optical transport in the weak-field regime, whereas quantum
mechanical methods become essential in the ultra-quantum limit.\\


\onecolumngrid\
\begin{center}\
\begin{figure}
\begin{tabular}{ccc}
\includegraphics[width=0.35\linewidth]{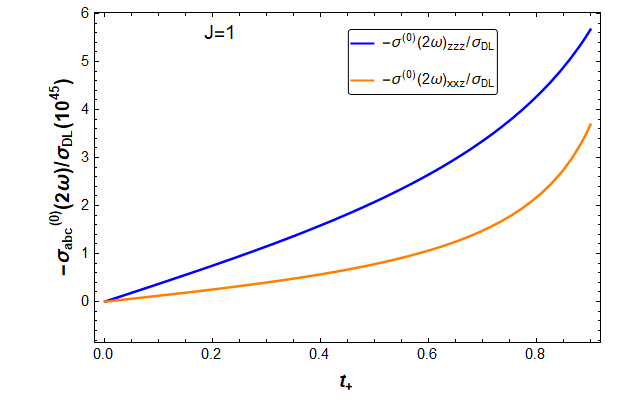} &
\includegraphics[width=0.35\linewidth]{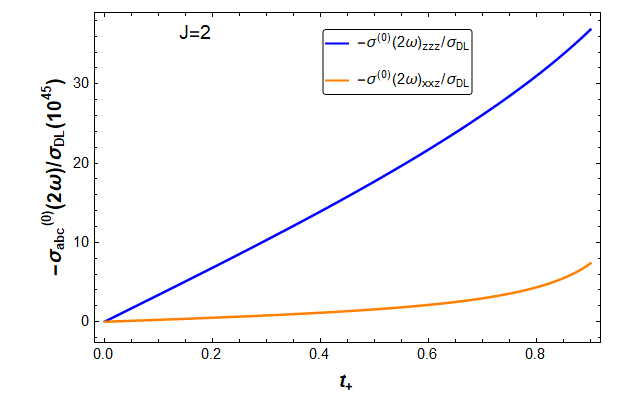} &
\includegraphics[width=0.35\linewidth]{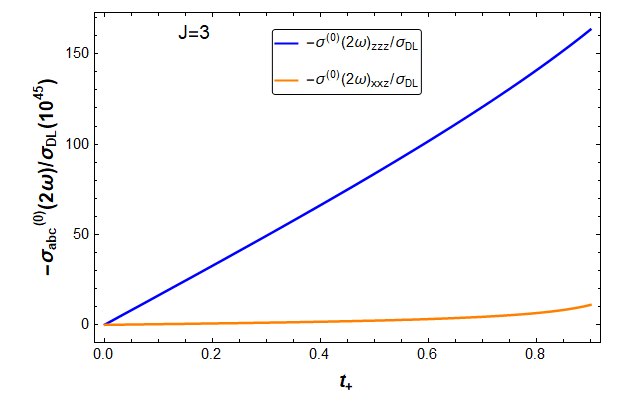} \\ 
\multicolumn{3}{c}{(a)}\\
\includegraphics[width=0.35\linewidth]{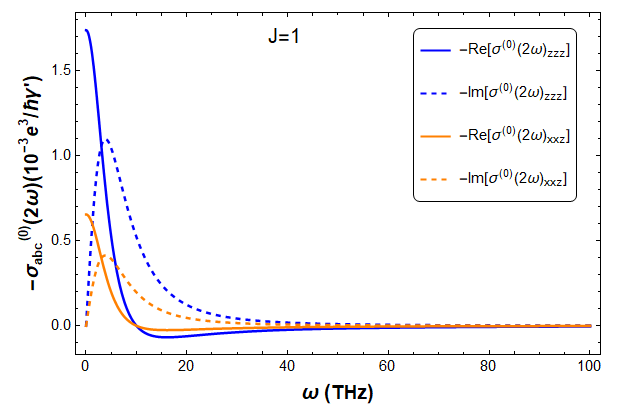} &
\includegraphics[width=0.35\linewidth]{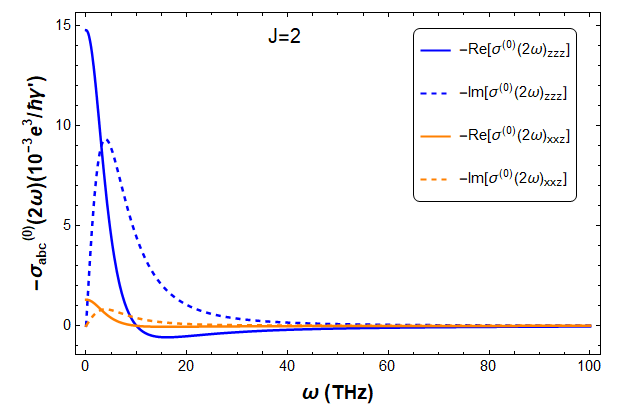} &
\includegraphics[width=0.35\linewidth]{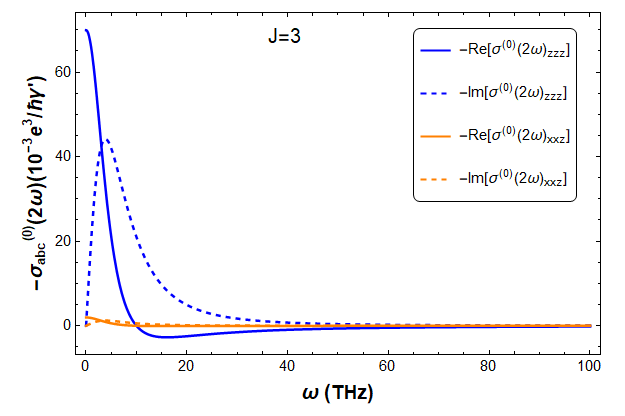} \\
\multicolumn{3}{c}{(b)}\\
\end{tabular}
\caption{(a) The tilt  $t_+$ dependence of the nonlinear conductivities $\sigma_{zzz}^0$ and $\sigma_{xxz}^0$. (b) The frequency dependence of conductivities $\sigma_{zzz}^0$ and $\sigma_{xxz}^0$  for the process of second harmonic generation at $t_s=0.5$. Here the relaxation rate $\gamma^{'}=\hbar/\tau$. All the other parameters are identical to those in Fig.(\ref{fig_cond_noB_tilt}).} \
\label{fig_cond_noB_second_har}\
\end{figure}\
\end{center}\
\twocolumngrid\

\section{Second order non-linear response of multi-WSMs}\label{non-linear}
We investigate the second-order nonlinear magneto-optical response of multi-Weyl semimetals. Substituting  Eq.(\ref{eomdecoupleb}) into Eq.(\ref{SBE}), ignoring the Lorentz force term, and keeping terms up to second order in E, we get

\begin{eqnarray}
\frac{1}{\hbar D}\Bigl[-e \textbf{E}-\frac{e^2}{\hbar}(\textbf{E}\cdot\textbf{B})\bm\Omega_{\bm k}^s\Bigr]\cdot \frac{\partial \tilde{f}_1^s}{\partial \bm k} -i \omega \tilde{f}_2^s=-\frac{\tilde{f}_2^s}{\tau}
\end{eqnarray}

Solve for $ \tilde{f}_2^s $, we obtain
\begin{equation}\label{eqf2}
\tilde{f}_2^s=\frac{\tau}{(1-2i\omega \tau)}\frac{1}{\hbar D}[e \textbf{E}+\frac{e^2}{\hbar}(\textbf{E}\cdot\textbf{B})\bm\Omega_{\bm k}^s]\cdot \frac{\partial \tilde{f}_1^s}{\partial \bm k}
\end{equation}

Substituting Eq.(\ref{linearE}) into Eq.(\ref{eqf2}) and retaining terms up to first order in $ \textbf{B} $, we obtain

\begin{widetext}
\begin{eqnarray}\label{eqnf2}
\tilde{f}_2^s&=&\frac{e\tau^2}{\hbar(1-i\omega \tau)(1-2i\omega \tau)}\biggl\lbrace\textbf{E}\cdot \frac{\partial}{\partial \bm k}\biggr[\Bigl(e \textbf{E}+\frac{e^2}{\hbar}(\textbf{E}\cdot \textbf{B})\bm\Omega_{\bm k}^s-\frac{e^2}{\hbar}(\textbf{B}\cdot \bm\Omega_{\bm k}^s)\textbf{E}\Bigr)\cdot \bm{\mathit{v}}_{\bm k}^s \frac{\partial f_0^s}{\partial \epsilon_{\bm k}^s}-\frac{e}{\hbar}\textbf{E}\cdot\frac{\partial}{\partial \bm k}\biggl(\textbf{m}_{\bm k}^s\cdot \textbf{B}\frac{\partial f_0^s}{\partial \epsilon_{\bm k}^s}\biggr)\biggr]\nonumber\\
&+&\biggl[\frac{e^2}{\hbar}(\textbf{E}\cdot \textbf{B})\bm\Omega_{\bm k}^s -\frac{e^2}{\hbar}(\textbf{B}\cdot \bm\Omega_{\bm k}^s)\textbf{E}\biggr]\cdot\frac{\partial}{\partial \bm k}\biggl(\textbf{E}\cdot \bm{\mathit{v}}_{\bm k}^s \frac{\partial f_0^s}{\partial \epsilon_{\bm k}^s}\biggr)\biggr\rbrace
\end{eqnarray}
\end{widetext}

Now the electric current density at the frequency $ 2\omega $ is given by

\begin{eqnarray}\label{cur_den_2}
\bm{j}_1&=&-\frac{e}{(2\pi)^3}\int d^3k \Bigl[\bm{\mathit{\tilde{v}}}_{\bm k}^s+\frac{e}{\hbar}(\bm\Omega_{\bm k}^s \cdot \bm{\mathit{\tilde{v}}}_{\bm k}^s )\textbf{B}\Bigr]\tilde{f}_2^{s} \nonumber\\
&-&\frac{e^2}{2\pi)^3\hbar}\int d^3k \textbf{E}\times \bm\Omega_{\bm k}^s \tilde{f}_1^s
\end{eqnarray}

According to the definition of second harmonic conductivity, this equation should be written in the frequency space $\omega$ as

\begin{equation}
\textbf{j}(2\omega)=\sigma(2\omega)\textbf{E}(\omega)\textbf{E}(\omega)
\end{equation}
where $\sigma(2\omega)  $ is the second harmonic conductivity.

\subsection{Second harmonic conductivity $\sigma_{abc}^{0}$ in the absence of magnetic field}
This subsection calculates the second harmonic current of the multi-Weyl semimetals system in the absence of magnetic fields (B = 0). A numerical analysis of the non-linear Hall effect in the tight-binding Hamiltonian of multi-Weyl semimetals can be found in Ref.\cite{roy2022non}.  The second harmonic conductivity tensor can be expressed as follows by inserting Eq.(\ref{eqnf2})  into the first term of Eq.(\ref{cur_den_2}).

\begin{equation}
\sigma_{abc}^0(2\omega)
=\frac{-\tau^{2}}{(1-i\omega \tau)(1-2i\omega \tau)}\frac{e^3}{\hbar(2\pi)^3}\int d^3k \frac{\partial \mathit{v}_a^s}{\partial k_c}  \mathit{v}_b^s
\Bigl(-\frac{\partial f_0^s}{\partial \epsilon_{\bm k}^s}\Bigr)
\end{equation}

Except for the aaz, aza, and zaa (a = x, y, z) components of the second harmonic conductivity tensor are nonzero and all other components are equal to zero \cite{gao2022suppression,gao2021second}.\\

\begin{widetext}
\begin{equation}
\begin{split}
\sigma_{zzz}^0(2\omega) = &-\frac{ \mathit{v_F^2} \alpha_J^{-2/J} \mu^{-1+\frac{2}{J}} \hbar^{-2/J}\sigma_{DL}}{ J\, \pi^{1/2}\, \Gamma\!\left(\frac{3}{2} + \frac{1}{J}\right)} (1 - t_s)^{-2/J} \Gamma\!\left(1 + \frac{1}{J}\right) \Bigg[ (-1 + t_s)\, {}_2F_1\!\left(1 + \frac{1}{J}, \frac{2}{J}; 2 + \frac{2}{J}; \frac{2t_s}{-1+t_s}\right) \\
&\quad + {}_2F_1\!\left(2 + \frac{1}{J}, \frac{2}{J}; 3 + \frac{2}{J}; \frac{2t_s}{-1+t_s}\right) \Bigg]
\end{split}
\end{equation}
\end{widetext}
\begin{equation}
\sigma_{xxz}^0(2\omega)=\frac{J\mu\sigma_{DL}}{\pi\hbar^2}\Bigl[\frac{3}{t_s^3}+\frac{3-t_s^2}{2t_s^4}\ln\frac{1-t_s}{1+t_s}\Bigr]
\end{equation}

\noindent where $\sigma_{DL}=\frac{e^3\tau^2}{4\pi\hbar(1-2i\omega \tau)(1-i\omega \tau)}$ is the Drude-like frequency dependent complex conductivity. Therefore, the above components follow the power laws $\sigma_{zzz}^{(0)}(2\omega) \propto \frac{ \mathit{v_F^2} \alpha_J^{-2/J} \mu^{-1+\frac{2}{J}}}{ J}$  and  $\sigma_{xxz}^{(0)} (2\omega)\propto J \mu $. The second harmonic conductivity tensor satisfies the relation $\sigma_{zxx}^0(2\omega)
=\sigma_{zyy}^0(2\omega)=\sigma_{xzx}^0(2\omega)=\sigma_{yzy}^0(2\omega)=\sigma_{xxz}^0(2\omega)=\sigma_{yyz}^0(2\omega)$ and it does not depend on the chirality of the Weyl node. The total second harmonic conductivity in tilted multi-Weyl semimetals is the sum of a pair of Weyl nodes. For the case with tilt inversion symmetry $t_{+} = -t_{-} , \sigma_{abc}(2\omega)= 0 $.  For the case with broken tilt inversion symmetry $t_{+} = t_{-} , \sigma_{abc}(2\omega)\neq 0$.\\

It is noted to see from Fig.(\ref{fig_cond_noB_second_har}a)  that the $\sigma_{abc}^0(2\omega)$ becomes exactly zero when $t_s \rightarrow 0$. The presence of the finite tilt is needed to get the second harmonic generation in mWSMs. The  $\sigma_{zzz}^0(2\omega)$  is more sensitive to tilt than $\sigma_{xxz}^0(2\omega)$ and their differences increase with J. Figure (\ref{fig_cond_noB_second_har}b) shows the frequency dependence of optical conductivities in multi-WSMs. Here, the real parts of the  $\sigma_{abc}^0(2\omega)$ exhibit positive and negative amplitudes due to the complex denominator function $\sigma_{DL}$ contribution to the second harmonic generation processes. Adopting a laser frequency $\omega \sim 2$ THz, the electric field strength $\sim 10^4 V/m$, and a laser spot size of 50-$\mu$m diameter in a mWSM, we obtain the nonlinear current in Table 1.  The nonlinear optical polarizability $\chi(0)(2\omega)$ for the process of the second harmonic generation is given by $\chi(0)(2\omega) = \sigma(0) (2\omega)/2 i \omega \epsilon_0$ , where $\epsilon_0$ is vacuum permittivity and tabulated $\chi(0)(2\omega)$ in Table 1. \\

\subsection{Second harmonic conductivity $\sigma_{abc}^{B}$ in the presence  of linear magnetic field}
Inserting Eq. (\ref{eqnf2}) into the first term of Eq. (\ref{cur_den_2}), we write the linear $ \textbf{B} $ dependent second harmonic conductivity tensor 
\begin{widetext}
\begin{eqnarray}\label{second_B_cond}
\sigma_{abc}^{(B)}(2\omega)&=&\frac{\tau^{2}}{(1-i\omega \tau)(1-2i\omega \tau)}\frac{e^3}{\hbar(2\pi)^3}\int d^3k \biggl\lbrace \frac{\partial \mathit{v}_a^s}{\partial k_c}\Bigl[-\frac{e B_b}{\hbar}(\bm\Omega_{\bm k}^s \cdot \bm{\mathit{v}}_{\bm k}^s )+\frac{e \mathit{v}_b^s}{\hbar}(\bm\Omega_{\bm k}^s \cdot \bm{B})\Bigr]-\frac{e B_c}{\hbar}\frac{\partial}{\partial \bm k}\cdot(\mathit{v}_a^s \bm\Omega_{\bm k}^s )\mathit{v}_b^s
\nonumber\\
&+&\frac{e}{\hbar} \frac{\partial [\mathit{v}_a^s(\bm\Omega_{\bm k}^s \cdot \bm{B})]}{\partial k_c}\mathit{v}_b^s-\frac{e B_a}{\hbar}\frac{\partial (\bm\Omega_{\bm k}^s \cdot   \bm{\mathit{v}}_{\bm k}^s )}{\partial k_c}\mathit{v}_b^s +\frac{\partial^2(\textbf{m}_{\bm k}^s\cdot \textbf{B})}{\hbar  \partial k_a \partial k_c}\mathit{v}_b^s -\frac{\partial^2 \mathit{v}_a^s}{\hbar  \partial k_a \partial k_c} (\textbf{m}_{\bm k}^s\cdot \textbf{B})\biggr\rbrace \Bigl(-\frac{\partial f_0^s}{\partial \epsilon_{\bm k}^s}\Bigr)\end{eqnarray}
\end{widetext}

We will explore Eq.(\ref{second_B_cond}) with following two cases:\\

Case-I In the case of $\textbf{B}\Vert \hat{\bm t}_s \Vert \hat{\bm z} $, for a single multi- Weyl node, we obtain
the magnetoconductivity components

\begin{widetext}
\begin{eqnarray}
\sigma_{xzx}^{(B)}(2\omega) &= &s\sigma_{2}^{(B)}\frac{ 2\alpha_J^{2/J} \pi^{1/2} (1 -t_s)^{2/J} \Gamma(2-\frac{1}{J}) }{3 \Gamma\!\left(\frac{3}{2} - \frac{1}{J}\right)\hbar^{3-\frac{2}{J}}\mu^{2(1/J-1)}} 
 \Bigg[ \frac{ \Big(4 + J\big(12J + 4J t_s
 + (4 - 5J)t_s^2 + 3(-1+J)J t_s^3 - 2(7+t_s)\big)\Big)}{(1-t_s)^{1/J}t_s^3 (1 + t_s)^{1-\frac{1}{J}} }\nonumber\\ &-&3(J-1)J^2 \frac{(1 - t_s)^{-1/J}} {(1 + t_s)^{1-\frac{1}{J}} }
- \frac{(2 - 3J)\big(2 + J(-4 + 3t_s^2)\big)\, {}_2F_1\!\left(2 - \frac{2}{J}, 2 - \frac{1}{J}; 3 - \frac{2}{J}; \frac{2t_s}{-1+t_s}\right)}{(-1 + t_s)^2 t_s^3} \Bigg]
\end{eqnarray}

\begin{equation}
\begin{split}
\sigma_{zxx}^{(B)}(2\omega) =s\sigma_{2}^{(B)} &\frac{ 2\pi^{1/2}\alpha_J^{2/J} \mu^{\frac{2J-2}{J}} \hbar^{-3+\frac{2}{J}}}{3 t_s^3\Gamma\!\left(\frac{3}{2} - \frac{1}{J}\right)} (1 - t_s)^{2/J} \Gamma(2-1/J)\Bigg[\frac{(1 + t_s)^{-1+\frac{1}{J}}}{(1-t_s)^{1/J}} \Big(2 - J\big(7 + t_s - 2t_s^2 + J(-6 + (-2+t_s)t_s)\big)\Big)\\&\quad  + \frac{1}{(-1 + t_s)^2} \big(-2 + J\big(7 - 6J + 3(-1+J)t_s^2\big)\big)\,{}_2F_1\!\left(2 - \frac{2}{J}, 2 - \frac{1}{J}; 3 - \frac{2}{J}; \frac{2t_s}{-1+t_s}\right) \Bigg]
\end{split}
\end{equation}
\end{widetext}

\noindent where $\sigma_{2}^{(B)}=\frac{e^4\tau^2 B}{(1-2i\omega \tau)(1-i\omega\tau)8\pi^2\hbar\mu}$. Therefore, the above components obey the power laws $\sigma_{xzx}^{(B)}(2\omega), \sigma_{zxx}^{(B)} (2\omega)\propto \alpha_J^{2/J} \mu^{2(1-1/J)} $. We can check that other nonzero second harmonic conductivity components satisfy the relations: $ \sigma_{zxx}^B(2\omega)=\sigma_{zyy}^B(2\omega)=\sigma_{xzx}^B(2\omega)=\sigma_{yzy}^B(2\omega) $. The J = 1 case is independent of tilt, but $J > 1$ depends on the even function of $t_s$. The aforementioned equations demonstrate the B-linear contribution to the second harmonic conductivity depend on chirality. Hence, when the conductivity over the Weyl cones with opposing chirality is combined, the total B-linear contribution to the second harmonic conductivity disappears.\\

Case-II In the case of $\textbf{B}\perp  \hat{\bm t}_s (\Vert \hat{\bm z})$, we get the nonzero conductivity
components

\begin{eqnarray}
\sigma_{zxz}^{(B)}(2\omega)&=& s\frac{2J\mathit{v}_F^2}{3} \sigma_2^{(B)}\cos\gamma\\
\sigma_{xzz}^{(B)}(2\omega)&=&- s\frac{2J\mathit{v}_F^2}{3}  \sigma_2^{(B)}\cos\gamma\\
\sigma_{zyz}^{(B)}(2\omega)&=& s\frac{2J\mathit{v}_F^2}{3}  \sigma_2^{(B)}\sin\gamma\\
\sigma_{yzz}^{(B)}(2\omega)&=&- s\frac{2J\mathit{v}_F^2}{3}  \sigma_2^{(B)}\sin\gamma
\end{eqnarray}

Hence, the above components follow the power laws $\sigma_{zxz}^{(B)}(2\omega)$, $\sigma_{zyz}^{(B)} (2\omega)\propto J$ , $\sigma_{xzz}^{(B)} (2\omega),\sigma_{yzz}^{(B)} (2\omega)\propto J$. In this instance, a single node's B-linear contribution to the second harmonic conductivity depends on chirality but is unaffected by tilt. The overall B-linear contribution to the second harmonic conductivity vanishes when the conductivity over the Weyl cones with opposing chirality is added together.\\

\subsection{The second order nonlinear Hall conductivity $\sigma_{abc}^{(H,0)}$ in the absence of magnetic field}
In this subsection, we study the second-order nonlinear Hall effect of multi-Weyl semimetals without a magnetic field. Inserting Eq.(\ref{disf}) into the second term in Eq. (\ref{cur_den_2}) and taking B = 0, we obtain the nonlinear Hall conductivity.

\begin{equation}
\sigma_{abc}^{(H,0)}=\sigma_{abc}^{(H,2\omega)}=\epsilon_{acd}\frac{e^3 \tau}{\hbar^2(1-i\omega\tau)}D_{bd}
\end{equation}

\noindent where  $ \epsilon_{acd} $ is the three-dimensional Levi-Civita antisymmetric tensor and $D_{bd}$ is called Berry curvature dipole(BCD). It is a first-order moment of the Berry curvature\cite{sodemann2015quantum, zeng2021nonlinear}.\\

One can calculate the non-zero components of the Berry curvature dipole 

\begin{equation}
D_{bd}=\frac{\hbar}{(2\pi)^3}\int d^3k \Omega_{d}^{s}\mathit{v}_{b}^{s}\Bigl(-\frac{\partial f_0^s}{\partial \epsilon_{\bm k}^s}\Bigr)
\end{equation}

\begin{eqnarray}
D_{xx}=D_{yy}&=&-s\frac{J}{16\pi^2}\Bigl[\frac{2}{t_s^2}+\frac{1-t_s^2}{t_s^3}\ln\frac{1-t_s}{1+t_s}\Bigr]\\
D_{zz}&=&-s\frac{J}{8\pi^2}\frac{t_s^2-1}{t_s^3}\Bigl[2t_s+\ln\frac{1-t_s}{1+t_s}\Bigr]  
\end{eqnarray}

One can notice that Berry curvature dipole components $D_{xx}(=D_{yy}),D_{zz} \propto J$. The BCD elements depend on chirality and even function of $t_s$. As a result, contributions from two Weyl nodes with opposing chirality precisely cancel each other out \cite{gao2022suppression, sodemann2015quantum, gao2020second}.
Surprisingly, the trace of BCD for a Weyl node is a universal quantity specified by its integer-valued monopole strength, independent of the details of its Hamiltonian \cite{armitage2018weyl, matsyshyn2019nonlinear}.\\

\begin{equation}
Tr[D]= \frac{J}{4\pi^2}
\end{equation}\\

\subsection{Linear B-contribution to the second order nonlinear Hall conductivity $\sigma_{abc}^{(H,B)}$}
The second-order nonlinear Hall effect in a weak magnetic field is the subject of our current discussion. The complex nonlinear Hall conductivity is obtained by inserting Eq.(\ref{disf})  into the second term of Eq.(\ref{cur_den_2}) while keeping terms up to the first power of B.

\begin{equation}
\sigma_{abc}^{(H,B)}=\epsilon_{acd}\frac{e^3 \tau}{\hbar^2(1-i\omega\tau)}\bigl[D_{bd}^{\Omega}+D_{bd}^{m}\bigr]\label{cond_berrydip}
\end{equation}

where

\begin{equation}
D_{bd}^{\Omega}=\frac{e}{(2\pi)^3}\int d^3k \Omega_{d}^{s}[B_b(\bm\Omega_{\bm k}^s \cdot   \bm{\mathit{v}}_{\bm k}^s )-\mathit{v}_{b}^{s}(\bm\Omega_{\bm k}^s \cdot \bm{B})]\Bigl(-\frac{\partial f_0^s}{\partial \epsilon_{\bm k}^s}\Bigr)
\end{equation}

\begin{equation}
D_{bd}^{m}=\frac{1}{(2\pi)^3}\int d^3k \frac{\partial \Omega_{d}^{s}}{\partial k_{b}}(\bm m_{\bm k}^s \cdot \bm{B})\Bigl(-\frac{\partial f_0^s}{\partial \epsilon_{\bm k}^s}\Bigr)
\end{equation}

\noindent where $D_{bd}^{\Omega}$ and $D_{bd}^{m}$ are Berry curvature dipole contributions due to the Berry curvature and the orbital magnetic moment, respectively.\\

Again we will consider the following two cases.\\

Case-I In the case of $\textbf{B}\Vert \hat{\bm t}_s \Vert \hat{\bm z} $, one obtains the Berry curvature dipole components.

\begin{widetext}
\begin{eqnarray}
D_{xx}^{\Omega}=D_{yy}^{\Omega}&=&\frac{15D_2^{(B)} }{8} (-2 + J) J^2 \sqrt{\pi}\,t_s\, \alpha_J^{2/J} \mu^{2-\frac{2}{J}} \hbar^{-2+\frac{2}{J}} \Gamma\!\left(3 - \frac{1}{J}\right) {}_2\tilde{F}_1\!\left(\frac{3}{2} - \frac{1}{J}, \frac{-1+J}{J}; \frac{9}{2} - \frac{1}{J}; t_s^2\right)\\
D_{zz}^{\Omega}&=&\frac{-15D_2^{(B)}}{4} (-2 + J) J^2 \sqrt{\pi}\,t_s\, \alpha_J^{2/J} \mu^{2-\frac{2}{J}} \hbar^{-2+\frac{2}{J}} \Gamma\!\left(3 - \frac{1}{J}\right) {}_2\tilde{F}_1\!\left(\frac{3}{2} - \frac{1}{J}, \frac{-1+J}{J}; \frac{9}{2} - \frac{1}{J}; t_s^2\right)
\end{eqnarray}

\begin{equation}
\begin{split}
D_{xx}^{m}=D_{yy}^{m} = &\frac{15D_2^{(B)}}{16 \Gamma\!\left(\frac{9}{2} - \frac{1}{J}\right)} (-2 + J) J \sqrt{\pi}\, t_s\, \alpha_J^{2/J} \mu^{2-\frac{2}{J}} \hbar^{-2+\frac{2}{J}} \Gamma\!\left(2 - \frac{1}{J}\right)
 \Bigg[ (-2 + 7J)\, {}_2F_1\!\left(\frac{3}{2} - \frac{1}{J}, \frac{-1+J}{J}; \frac{7}{2} - \frac{1}{J}; t_s^2\right) \\
&\quad - 9J\, {}_3F_2\!\left(\frac{5}{2}, 1 - \frac{1}{J}, \frac{3}{2} - \frac{1}{J}; \frac{3}{2}, \frac{9}{2} - \frac{1}{J}; t_s^2\right) \Bigg]
\end{split}
\end{equation}

\begin{equation}
\begin{split}
D_{zz}^{m} = &\frac{15D_2^{(B)}}{16} (-2 + J) J^2 t_s\, \alpha_J^{2/J} \mu^{2-\frac{2}{J}} \hbar^{-2+\frac{2}{J}} \Gamma\!\left(2 - \frac{1}{J}\right) \Bigg[ -4 \sqrt{\pi}\, {}_2\tilde{F}_1\!\left(\frac{3}{2} - \frac{1}{J}, \frac{-1+J}{J}; \frac{7}{2} - \frac{1}{J}; t_s^2\right) \\
&\quad + 9\pi\, {}_3\tilde{F}_2\left(\frac{5}{2}, \frac{-1+J}{J}, \frac{3}{2} - \frac{1}{J}; \frac{3}{2}, \frac{9}{2} - \frac{1}{J}; t_s^2\right) \Bigg]
\end{split}
\end{equation}
\end{widetext}

where, $D_2^{(B)}=\frac{e B \hbar}{120\pi^2 \mu^2}$.
  
Case-II In the case of $\textbf{B}\perp  \hat{\bm t}_s (\Vert \hat{\bm z})$ , we get the nonzero components

\begin{equation}
D_{zx}^{\Omega}=-6J t_s D_2^{(B)}\cos\gamma
\end{equation}

\begin{equation}
D_{zy}^{\Omega}=-6 J t_s D_2^{(B)}\sin\gamma
\end{equation}

\begin{widetext}
\begin{equation}
\begin{split}
D_{xz}^{\Omega} = &\frac{3J\sqrt{\pi}\, \alpha_J^{2/J} \mu^{2-\frac{2}{J}} \hbar^{-2+\frac{2}{J}} \Gamma\!\left(2 - \frac{1}{J}\right)}{4t_s} \Bigg[ (2 + J)\big(-5 + 2J + 10t_s^2\big)\, {}_2\tilde{F}_1\!\left(\frac{1}{2} - \frac{1}{J}, \frac{-1+J}{J}; \frac{7}{2} - \frac{1}{J}; t_s^2\right) \\
&\quad + (-2 + J)(-1 + 2J)\big(-1 + t_s^2\big)\, {}_2\tilde{F}_1\!\left(\frac{3}{2} - \frac{1}{J}, \frac{-1+J}{J}; \frac{7}{2} - \frac{1}{J}; t_s^2\right) \Bigg]\cos\gamma
\end{split}
\end{equation}

\begin{equation}
\begin{split}
D_{yz}^{\Omega} = &\frac{3J\sqrt{\pi}\, \alpha_J^{2/J} \mu^{2-\frac{2}{J}} \hbar^{-2+\frac{2}{J}} \Gamma\!\left(2 - \frac{1}{J}\right)}{4t_s} \Bigg[ (2 + J)\big(-5 + 2J + 10t_s^2\big)\, {}_2\tilde{F}_1\!\left(\frac{1}{2} - \frac{1}{J}, \frac{-1+J}{J}; \frac{7}{2} - \frac{1}{J}; t_s^2\right) \\
&\quad + (-2 + J)(-1 + 2J)\big(-1 + t_s^2\big)\, {}_2\tilde{F}_1\!\left(\frac{3}{2} - \frac{1}{J}, \frac{-1+J}{J}; \frac{7}{2} - \frac{1}{J}; t_s^2\right) \Bigg]\sin\gamma
\end{split}
\end{equation}
\end{widetext}

and
\begin{eqnarray}
D_{zx}^{m}&=&-3Jt_s D_2^{(B)}\cos\gamma\\
D_{zy}^{m}&=&-3Jt_s D_2^{(B)}\sin\gamma
\end{eqnarray}

Therefore, the above components follow the power laws $D_{zx}^{\Omega},D_{zx}^{m}\propto J$, $D_{xz}^{\Omega},D_{xz}^{m}  \propto \alpha_J^{2/J}\mu^{2(1-\frac{1}{J})}$ , $D_{zy}^{\Omega},D_{zy}^{m} \propto J $ and $ D_{yz}^{\Omega} ,D_{yz}^{m}\propto \alpha_J^{2/J}\mu^{2(1-\frac{1}{J})}$. One can check that $D_{zx}^{m} =D_{xz}^{m} $ and $D_{zy}^{m}=D_{yz}^{m}  $ and rest of the components are zero. \\

The nonlinear Hall effect can be modulated by the polarization of the incident light, as discussed in Ref.\cite{gao2022suppression}. Using Eq. (\ref{cond_berrydip}), the electric current is rewritten in the form of

\begin{equation}\label{current_density}
\textbf{j}(2\omega)=\frac{e^3\tau}{\hbar^2(1-i\omega \tau)}(\hat{D}\cdot \textbf{E})\times \textbf{E}
\end{equation}

\onecolumngrid\
\begin{center}\
\begin{figure}
\begin{tabular}{ccc}
\includegraphics[width=0.5\linewidth]{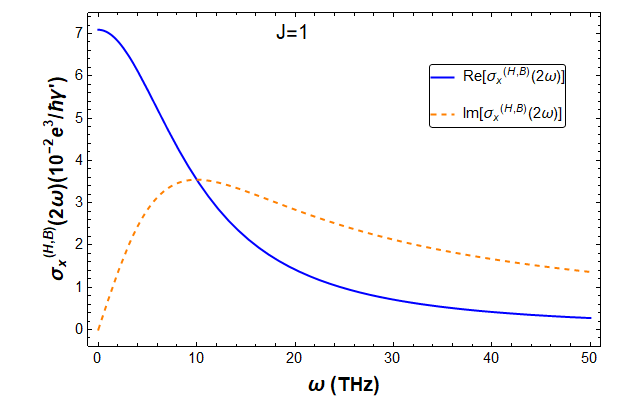} &
\includegraphics[width=0.5\linewidth]{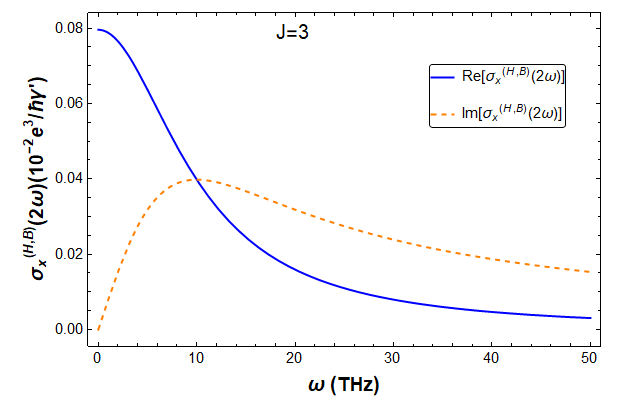} \\ 
\multicolumn{2}{c}{(a)}\\
\includegraphics[width=0.42\linewidth]{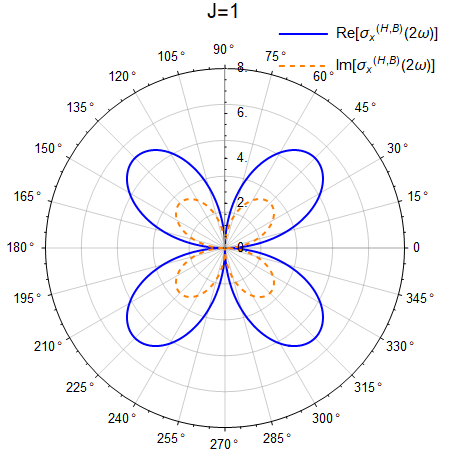} &
\includegraphics[width=0.48\linewidth]{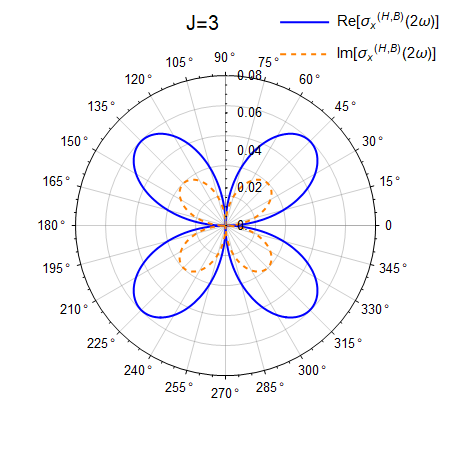} \\
\multicolumn{2}{c}{(b)}\\
\end{tabular}
\caption{(a) The nonlinear Hall conductivities $\sigma_{x}^{(H,B)}$ for the process of second harmonic generation as a function of the incident photon frequency on the tilt at $t_s=0.5$, B=1 T and $\gamma=\pi/4$. (b) The angle dependence of the nonlinear Hall conductivity $\sigma_{x}^{(H,B)}$ at $\omega=5$ THz. Here the relaxation rate $\gamma^{'}=\hbar/\tau$. All the other parameters are identical to those in Fig.(\ref{fig_cond_noB_tilt}).}  \
\label{fig_cond_nonlinear_hall}\
\end{figure}\
\end{center}\
\twocolumngrid\
Assume that an electromagnetic wave propagates in the x direction:
\begin{equation}\label{e_field}
\textbf{E}(\textbf{r},t)=\vert\textbf{E}(\omega)\vert Re[\vert \psi\rangle e^{i(q x- \omega t)}
\end{equation}

where
\begin{equation}
\vert \psi\rangle=\Bigl(
\begin{array}{c}
\psi_y\\
\psi_z
\end{array}\Bigr)=\Bigl(
\begin{array}{c}
\sin \gamma e^{i \alpha_y}\\
\cos \gamma e^{i \alpha_z}
\end{array}\Bigr)
\end{equation}

\noindent is the Jones vector in the y-z plane with phases $ \alpha_y, \alpha_z$  and
the amplitudes $E_y = \vert \textbf{E}\vert \sin \gamma$, $E_z = \vert \textbf{E}\vert \cos \gamma$. Inserting Eq. (\ref{e_field}) into Eq. (\ref{current_density}), we obtain the nonlinear Hall current

\begin{equation}\label{current_x}
j_{x}(2\omega)=\frac{e^3\tau}{\hbar^2(1-i\omega \tau)}\frac{D_{yy}^{(B)}-D_{zz}^{(B)}}{2}\sin 2\gamma e^{(\alpha_y+\alpha_z)}\vert\textbf{E}\vert^{2}
\end{equation}

Following the same procedure of Ref.(\cite{gao2022suppression}), Eq.(\ref{current_x}) can be rewritten as $ j_x \sim (D_{yy}-D_{zz})E_y E_z$ . Hence conductivity will be

\begin{widetext}
\begin{equation}
\begin{split}
\sigma_{x}^{(H,B)}(2\omega)= &\frac{e^3\tau}{(1-i\omega\tau)\hbar^2}\frac{9}{8t_s} (-2 + J) \sqrt{\pi}\alpha_J^{2/J} \mu^{2-\frac{2}{J}} \hbar^{-2+\frac{2}{J}} \Gamma\!\left(2 - \frac{1}{J}\right)\Bigg[ (-2 + J)\big(7 - 6J + 5(-1 + J)t_s^2\big)\, \\
& {}_2\tilde{F}_1\!\left(\frac{3}{2} - \frac{1}{J}, \frac{-1+J}{J}; \frac{7}{2} - \frac{1}{J}; t_s^2\right)- \big(2 + J(-7 + 6J)\big)\big(-1 + t_s^2\big)\, {}_2\tilde{F}_1\!\left(\frac{5}{2} - \frac{1}{J}, \frac{-1+J}{J}; \frac{7}{2} - \frac{1}{J}; t_s^2\right) \Bigg]{D_2^{(B)}\sin2\gamma}
\end{split}
\end{equation}
\end{widetext}

The linear-B second harmonic conductivities $\sigma_{x}^{(H,B)}(2\omega)$ as a function of incident photon frequency $\omega$ is depicted in Fig.(\ref{fig_cond_nonlinear_hall}a). This component has zero value for double WSM, in contrast to single and triple WSMs. The B-linear contribution to the nonlinear Hall conductivity is independent of the chirality and the odd function of $t_s$, in contrast to the situation when B = 0. The conductivity $\sigma_{x}^{(H,B)}(2\omega) \neq 0$ only occurs in the system with broken tilt inversion symmetry ($t_{+} = t_{-} $). As further shown in Fig.(\ref{fig_cond_nonlinear_hall}b), it is clear from Eq. (\ref{current_x}) that $\sigma_{x}^{(H,B)}(2\omega)$  achieves its maximum when the polarization direction $\gamma= \pm \pi/4$ and vanishes at $\gamma= 0, \pi/2$.\\

\begin{table*}
\caption{\label{tab:table3}The comparison among the conductivity elements of multi-Weyl semimetals. The other parameters are taken as $v_F=5\times 10^5$ m / s, $\alpha_2=0.009 m^2/s$, $\alpha_3=4.5\times 10^{-11} m^3/s$, $\mu=1 meV$ and $\tau=10^{-13}s$, B=1T and $t_{+}$=0.5.}
\begin{ruledtabular}
\begin{tabular}{ccccc}
Conductivity &J=1&J=2	&J=3\\ \hline
$\sigma_{xx}^{(0)}$ & 3.26 $\Omega^{-1} m^{-1}$ & 6.51 $\Omega^{-1} m^{-1}$ & 9.77 $\Omega^{-1} m^{-1}$ \\
$\sigma_{zz}^{(0)}$
& 2.74 $\Omega^{-1} m^{-1}$ & 36.32 $\Omega^{-1} m^{-1}$ & 486.79 $\Omega^{-1} m^{-1}$\\
 $\sigma_{xx}^{(B)}$  & 76.59 $\Omega^{-1} m^{-1}$ & 32.52 $\Omega^{-1} m^{-1}$ & 6.00 $\Omega^{-1} m^{-1}$ \\
 $\sigma_{zz}^{(B)}$  & 194.17 $\Omega^{-1} m^{-1}$ & 388.35 $\Omega^{-1} m^{-1}$ & 582.52$\Omega^{-1} m^{-1}$ \\
$\rho_{xy}$  & 21.26 $\times 10^{-3} \Omega$-m(at 2THz)  & 31.90  $\times 10^{-4} \Omega$-m(at 2THz) & 118.16  $\times 10^{-4} \Omega$-m(at 2THz)\\
 $i_{zzz}$  & 9.65$\times 10^{-6} A $ & 8.22 $\times 10^{-5} A $ & 8.08$\times 10^{-4} A $ \\
		 
 $\i_{xxz}$  & 3.63 $\times 10^{-6}$ A & 7.27 $\times 10^{-6} A$ & 1.09 $\times 10^{-5} A $\\
		 
 $\chi_{zzz}$  & 9.05 $\times 10^{5}$ & 7.71 $\times 10^{6}$ & 7.59 $\times 10^{7}$\\
 $\chi_{xxz}$  & 5.22 $\times 10^{5}$ & 1.04 $\times 10^{6}$ & 1.57 $\times 10^{6}$\\
 $i_x(H,B)$  & 4.97 $\times 10^{-4} A$  & 0 & 2.68 $\times 10^{-6} A$\\
 $\chi_x(H,B)$  & 1.43 $\times 10^{7} pm/V$  & 0 & 7.71 $\times 10^{5} pm/V$\\
\end{tabular}
\end{ruledtabular}
\end{table*}

\section{Conclusion}\label{conclusion}
The present work provides a unified semiclassical description of the linear and nonlinear
magneto-optical transport properties of tilted  multi-Weyl semimetals by
simultaneously incorporating the effects of Berry curvature (BC) and the orbital magnetic
moment (OMM). Although both quantities originate from the geometric properties of Bloch
wave functions, they influence transport in fundamentally different ways. The BC primarily generates anomalous carrier motion through the transverse velocity,
whereas the OMM modifies the quasiparticle energy spectrum and
consequently alters the group velocity and equilibrium carrier distribution. Their
competition governs the overall magneto-optical response of the system.\\

Another significant finding is the existence of universal scaling relations between the optical conductivity, the chemical potential, and the monopole charge. Throughout the manuscript, different conductivity tensors obey characteristic power laws. These scaling relations arise naturally from the anisotropic electronic dispersion of multi-Weyl
semimetals and provide experimentally measurable fingerprints of higher-order topological
fermions. Since the density of states increases with increasing monopole charge, optical
absorption and nonlinear responses become considerably stronger for double and triple
Weyl semimetals than for ordinary Weyl systems.\\

The present work also establishes the importance of the Berry-curvature dipole in governing
the nonlinear Hall effect and reveals that its trace is a universal quantity determined solely
by the monopole charge. Overall, our theoretical framework unifies the description of linear and nonlinear magneto-
optical transport in tilted multi-Weyl semimetals and provides experimentally testable
predictions for terahertz spectroscopy, nonlinear optical measurements and Hall transport experiments. We expect that the analytical results presented here will stimulate further
theoretical investigations and experimental studies of topological quantum materials with
higher monopole charge.\\

Finally, although the present analysis has been performed within the semiclassical
Boltzmann framework, the theoretical approach can be extended in several directions.
Possible future studies include finite-temperature transport, disorder-induced quantum
corrections, strong magnetic-field regimes involving Landau-level quantization, electron-
electron interaction effects, strain-induced pseudo-magnetic fields and nonlinear optical
responses in type-II multi-Weyl semimetals. Such extensions would further deepen our
understanding of magneto-optical transport in topological quantum materials.\\

\section{Acknowledgements}
We are grateful for helpful discussions with Debanand Sa. We have not received any financial support from any source for this research.

\begin{appendix}
\section{DETAILS OF THE CALCULATIONS USING SPHERICAL POLAR COORDINATES}
In this paper, we focus on the n-doped multi-Weyl semimetals with a positive chemical potential $\mu$. In general, one can decompose the momentum k into parallel and perpendicular parts:

\begin{eqnarray}
k_x&=& \Bigl(\frac{k \sin \theta}{\alpha_J}\Bigr)^{\frac{2}{J}}\cos \phi \\
k_x&=& \Bigl(\frac{k \sin \theta}{\alpha_J}\Bigr)^{\frac{2}{J}}\sin \phi \\
k_z&=& \frac{k}{v_F}\cos \theta
\end{eqnarray}

The Jacobian of the transformation is $\mathcal{J} =   \frac{1}{J v_F \sin \theta } \left( \frac{k \sin \theta}{\alpha_{J} } \right) ^{2/J} $, which has been used for analytical expressions of conductivity elements of multi-WSMs.
\end{appendix}

\bibliography{tilted_mWSMs}
\end{document}